REVIEW

# Nanomechanical Resonators: Toward Atomic Scale

Bo Xu,[†] Pengcheng Zhang,[†] Jiankai Zhu,[†] Zuheng Liu,[†] Alexander Eichler, Xu-Qian Zheng, Jaesung Lee, Aneesh Dash, Swapnil More, Song Wu, Yanan Wang, Hao Jia, Akshay Naik,* Adrian Bachtold, Rui Yang,* Philip X.-L. Feng,* and Zenghui Wang*

Cite This: ACS Nano 2022, 16, 15545−15585

Read Online

ACCESS | 📊 Metrics & More | 📄 Article Recommendations

**ABSTRACT:** The quest for realizing and manipulating ever smaller man-made movable structures and dynamical machines has spurred tremendous endeavors, led to important discoveries, and inspired researchers to venture to previously unexplored grounds. Scientific feats and technological milestones of miniaturization of mechanical structures have been widely accomplished by advances in machining and sculpturing ever shrinking features out of bulk materials such as silicon. With the flourishing multidisciplinary field of low-dimensional nanomaterials, including one-dimensional (1D) nanowires/nanotubes and two-dimensional (2D) atomic layers such as graphene/phosphorene, growing interests and sustained effort have been devoted to creating mechanical devices toward the ultimate limit of miniaturization—genuinely down to the molecular or even atomic scale. These ultrasmall movable structures, particularly nanomechanical resonators that exploit the vibratory motion in these 1D and 2D nano-to-atomic-scale structures, offer exceptional device-level attributes, such as ultralow mass, ultrawide frequency tuning range, broad dynamic range, and ultralow power consumption, thus holding strong promises for both fundamental studies and engineering applications. In this Review, we offer a comprehensive overview and summary of this vibrant field, present the state-of-the-art devices and evaluate their specifications and performance, outline important achievements, and postulate future directions for studying these miniscule yet intriguing molecular-scale machines.

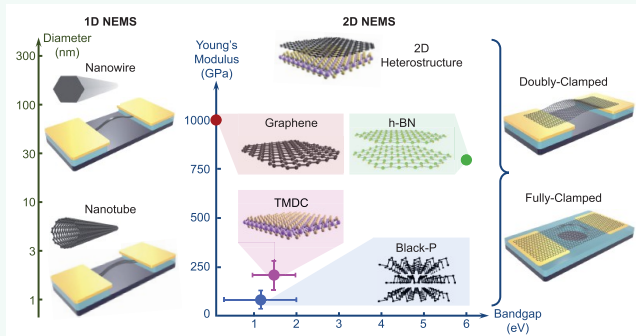

## 1. INTRODUCTION

Exploring mechanical degrees of freedom in genuinely nanoscale structures, such as molecules and nanomaterials, can lead to intriguing findings and devices at unconventional scales less familiar with everyday life experiences but imagination-capturing. For example, the exploration and demonstration of molecular rotors, shuttles, valves, muscles, transporters, etc., have witnessed the rise of the exciting research field of molecular machines, highlighted by the 2016 Nobel Prize in Chemistry.[1−3] Thanks to their infinitesimal sizes, molecular machines can engage in very fast linear/translational motion or rotary motion, for example, in molecular motors the rotational frequency can reach 12 megahertz (MHz), equivalent to millions of cycles per second.[1,4,5] Such exquisite and high-speed motions at the molecular scale are fascinating and suggest far-reaching potential for both fundamental research and future applications.

Among the different categories of nanoscale mechanical structures, nanoelectromechanical systems (NEMS) stand out as an important hallmark of devices, which leverage the mechanical degrees of freedom in nanostructures through deliberately designed device geometries. Further, by coupling different (e.g., electrical, optical, thermal, and magnetic) excitations (or information carriers) to the mechanical motion, NEMS structures have enabled a plethora of device research and applications.

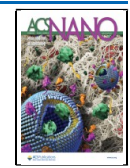

ACS Publications

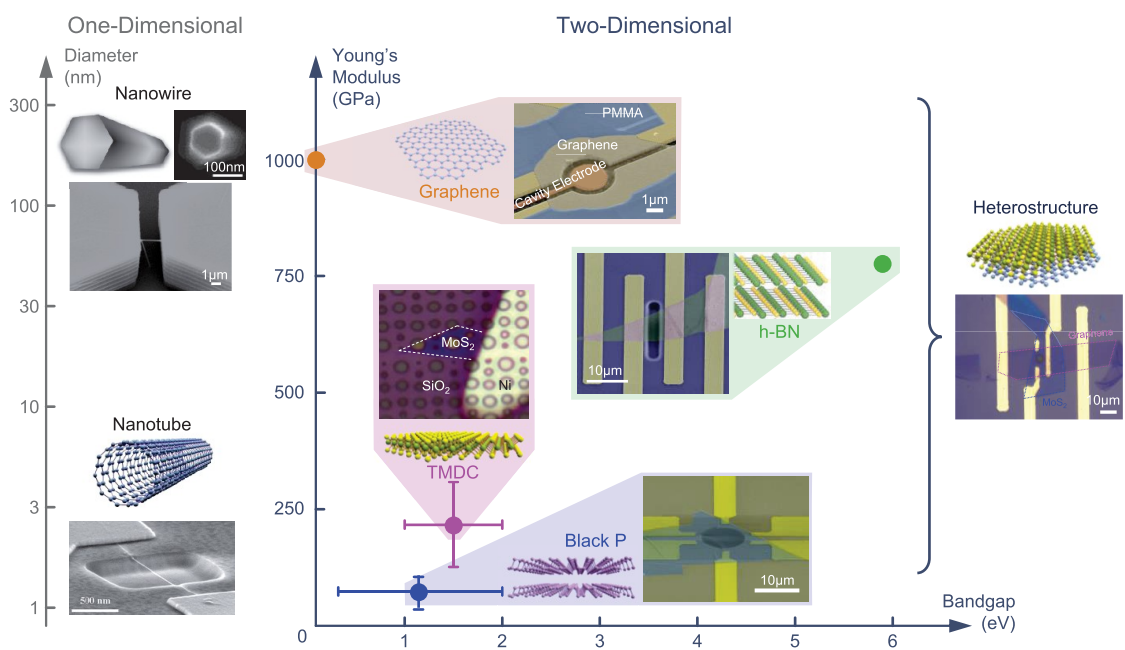

**Figure 1.1.** Nanomechanical resonators enabled by diverse 1D and 2D nanomaterials. SEM images and optical images are reprinted in part with permission from ref 10, copyright 2007 American Chemical Society; from ref 19, copyright 2015 Royal Society of Chemistry; from ref 63, copyright 2018 The Authors, some rights reserved; exclusive licensee AAAS. Distributed under a CC BY-NC 4.0 license http://creativecommons.org/licenses/by-nc/4.0/; from ref 125, copyright 2014 American Chemical Society; from ref 179, copyright 2021 American Chemical Society; from ref 198, copyright 2006 American Chemical Society; and under a Creative Commons (CC BY) License from ref 20, copyright 2017 Springer Nature, respectively.

Compared with quasi-static behaviors, dynamical operations of NEMS are more important and relevant. The most-studied type is NEMS resonators, which harness the vibrational modes in nanostructures, and have been demonstrated using different nanomaterials. One important category is derived from one-dimensional (1D) nanomaterials, such as nanowires (NWs) and nanotubes (NTs),[6−9] in which the extreme aspect ratio (length over diameter or width) and outstanding mechanical properties are leveraged to form high-performance nanomechanical resonators. For example, suspended silicon (Si) NWs grown over microtrenches can form doubly clamped NEMS resonators, with resonance frequency over 200 MHz.[10] Carbon nanotubes (CNTs) can also be used to build NEMS resonators.[11] With their low flexural stiffness and large stretchability, they can realize highly tunable resonant responses: In one example, it was possible to tune the resonance frequency of a device by more than 2000%.[12] Such broad tuning range is not readily attainable in mechanical resonators machined from bulk materials (such as microfabricated silicon resonators) or even resonators based on thicker Si NWs.

Suspending two-dimensional (2D) materials enables fabrication of atomically thin mechanical resonators. The exploration of 2D NEMS resonators started with graphene. Thanks to the material's high Young's modulus (up to 1 terapascal, TPa),[13] ultrasmall mass due to the atomic thickness (0.335 nm), and high carrier mobility (over $10^6$ cm$^2$ V$^{-1}$ s$^{-1}$ at low temperature),[14,15] graphene resonators have been under intensive research since 2007[16] and have led to many exciting findings. Over the time, NEMS resonators based on 2D materials other than graphene, such as transition-metal dichalcogenides (TMDCs),[17,18] black phosphorus (black P),[19] hexagonal boron nitride (h-BN),[20] 2D magnetic layers,[21,22] and wide-bandgap crystals[23] have also emerged, giving rise to a plethora of exciting research discoveries. Interestingly, 2D materials can be stacked into heterostructures (HSs), enabling nearly endless possibilities for creating and optimizing different types of 2D NEMS resonators.

In this Review, we offer a comprehensive and detailed survey about NEMS resonators based on 1D and 2D nanomaterials (Figure 1.1). We start by introducing the material choice and fabrication techniques (Section 2), followed by excitation and detection techniques (Section 3). We then summarize the state-of-the-art device specifications and performance in different metrics for both 1D (Section 4) and 2D NEMS resonators (Section 5). Section 6 discusses mode shapes (spatial profile of different resonant modes) and techniques for visualizing them. We then discuss the tuning of resonance frequency (Section 7) as well as quality ($Q$) factor and damping in NEMS resonators (Section 8). Dynamic response beyond simple harmonic motion, such as nonlinearity (Section 9) and mode coupling behavior (Section 10) are also presented. In terms of applications, we survey the demonstration of NEMS resonators in sensing (Section 11), signal processing, and fundamental researches such as quantum experiments (Section 12). We conclude by sharing our vision about the challenges and opportunities in this highly vibrant field (Section 13).

## 2. MATERIAL CHOICE AND FABRICATION TECHNIQUES

### 2.1. Low-Dimensional Materials Used for NEMS Resonators.
Many different types of 1D and 2D materials have been explored for realizing NEMS resonators, which demonstrate different mechanical, optical, magnetic, and electrical properties. The list has grown significantly over the years and is still expanding.

*2.1.1. 1D Nanomaterials for NEMS Resonators.* This category mostly includes NWs and NTs based on different materials. Among NWs,[24−29] Si NW is the most commonly explored type for building NEMS resonators, while CNT is by far the most extensively studied type of NTs in this regard. In fact, the exploration of low-dimensional material-based NEMS resonators largely started with CNT resonators,[11] which have been demonstrated by many research groups over the years, leading to a number of important findings and a series of records being refreshed.[30]

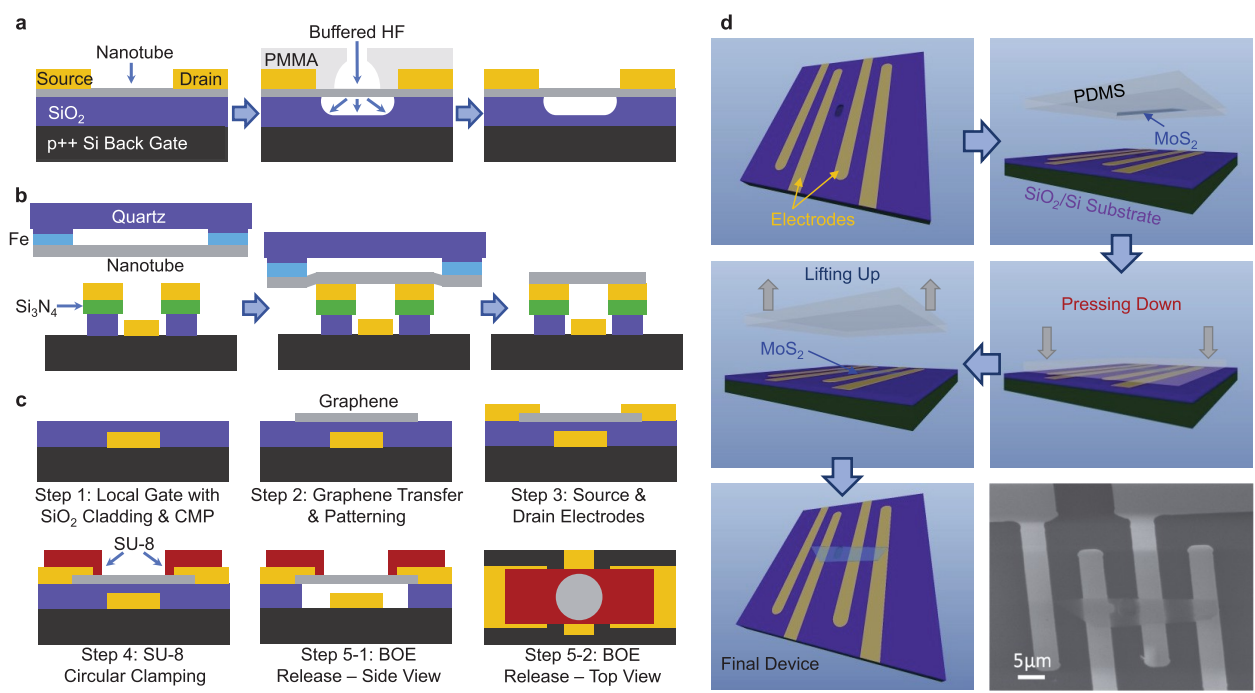

**Figure 2.1.** Schematic illustrations of fabrication processes for 1D and 2D NEMS resonators. (a) Suspending a nanotube by etching the sacrificial layer underneath using buffered hydrofluoric acid (HF) etching. Reprinted in part with permission from ref 198. Copyright 2013 American Chemical Society. (b) One-step direct transfer of CNTs for NEMS resonators and other functional devices. Reprinted in part with permission from ref 49. Copyright 2010 American Chemical Society. (c) Etching of sacrificial layer underneath graphene, with SU-8 circularly clamping the graphene to form a circular resonator with a local gate electrode. Reprinted in part with permission from ref 58. Copyright 2013 American Physical Society. (d) Dry transfer of molybdenum disulfide ($MoS_2$) onto a substrate with predefined microtrenches and contact electrodes. Reprinted in part with permission from ref 61. Copyright 2014 AVS American Institute of Physics.

*2.1.2. 2D Nanomaterials for NEMS Resonators.* Although started later than 1D NEMS, the research of 2D NEMS has been growing fast, and many different types of 2D materials have been explored for building NEMS resonators.[31] This list now includes graphene, 2D semiconductors, 2D magnets, wide-bandgap 2D materials, and continues to expand. Furthermore, HSs based on different combinations of 2D materials offer researchers almost unlimited possibilities to explore different types of NEMS resonators with desirable properties.

## 2.2. Fabrication Techniques Enabling NEMS Structures.
Various fabrication techniques, including both top-down and bottom-up approaches, have been developed and implemented in creating NEMS resonators. To date, most 1D and 2D devices are fabricated using a combination of top-down defined device geometries and bottom-up grown or synthesized nanomaterials.

*2.2.1. Fabrication Techniques for 1D NEMS Resonators.* With the advances of micro- and nanofabrication technologies, the microelectromechanical systems (MEMS) research community has been pushing the scaling of MEMS devices, and some 1D NEMS resonators have been built entirely using intricate top-down lithographic processes. For example, a silicon nanobeam resonator with the size of 7.7 $\mu$m × 300 nm × 800 nm has been fabricated using a dry etching process.[32] Similar examples include platinum (Pt) NW resonators[33] and rhodium (Rh) NW electromechanical arrays.[34] Such approaches are good for wafer-scale fabrication of NEMS devices; however, they are limited in material choices, have limits on their feature size (typically, the smallest device dimensions are still >100 nm), and could create surface defects during the processes which are detrimental to the device performance. Therefore, pure top-down approaches face great challenges in harnessing the material properties offered by low-dimensional material systems.

To address these challenges, also toward the goal of realizing higher-quality, atomically terminated surfaces and further miniaturized sizes, NEMS resonators based on bottom-up synthesized Si NWs have been developed, in which Si NWs are grown over predefined microtrenches using the vapor—liquid—solid (VLS) epitaxial growth,[35] forming Si NW NEMS resonators with a hexagonal cross section and a diameter of ~60 nm.[10] NEMS resonators based on tin dioxide ($SnO_2$) NWs have also been demonstrated by dry transferring the NWs from the growth substrate onto the target substrate with predefined contact electrodes.[36]

CNTs, which are chemically synthesized or grown, represent a molecular-level scaling limit (in diameter) for 1D NEMS resonators, with diameters down to the one nanometer level for single-walled nanotubes. Most CNTs used in NEMS resonators are grown in a chemical vapor deposition (CVD) process from gas such as methane ($CH_4$), ethane ($C_2H_6$), or acetylene ($C_2H_2$).[37] As carbon atoms dissociate from the gas, CNTs grow from catalyst islands typically containing iron and nickel. The initial nucleation determines the "chirality" of the CNTs, i.e., the angle of the tube axis relative to the honeycomb atomic lattice. Careful adjustment and monitoring of the growth parameters (temperature, gas composition, flow rate, etc.) enables controlling the nanotube radius and its structural quality to some extent.

Historically, the first method to create doubly clamped nanotube resonators was to grow CNTs on a substrate, on which microtrenches are subsequently patterned and etched (Figure 2.1a) after carefully locating and aligning with the selected CNTs during lithographical definition of the microtrenches.[11,38] Later, a method has been adopted to grow suspended nanotubes in the final fabrication step,[39] bridging two electrodes already separated by a microtrench.[40] This method has an advantage in that the nanotube never comes into contact with photo/e-beam resist, etchants, or cleaning solvents. The challenge is that all components of the chip, such as electrical leads, must withstand the rough CVD treatment at high temperatures. The CNTs grown this way have very clean surfaces (free from additional chemical processing), and such resonators have been used for surface science studies.[41−43] In this way, one can also create very long (mm level) CNT resonators.[44]

However, to date no reliable method exists to grow CNT resonators with the desired chirality and the controlled direction and length. This outstanding issue presents a serious obstacle for widespread applications of nanotubes as sensors (e.g. for adsorption detection). In addition, nanotubes exposed to air could accrue surface adsorbates, which could potentially affect device responses. Such adsorbates can be removed in situ by electrothermal annealing in ultrahigh vacuum.[45] By applying a bias voltage (a few volts) along the suspended nanotube, the dissipated electrical power heats the CNT sufficiently to remove adsorbates. Directly after the current annealing, the CNT can be clean enough to enable the adsorption of solid and fluid monolayers of noble gases.[46,47] Furthermore, current annealing also improves the mechanical quality factor of CNT resonators.[48] The removal of the surface species is more efficient in devices with low contact resistances, which is predominantly achieved by growing CNTs directly onto metal electrodes.[45]

Besides direct growth, alternative techniques have been developed by growing CNTs between the prongs of a nanoscale fork or similar structures.[49−52] Such structures can later be stamped onto a second substrate to transfer the nanotubes onto a chip[52] with electrodes (Figure 2.1b). This method combines the advantages of retaining clean nanotubes and creating complex electronic circuits.[53]

It is worth mentioning that while most 1D resonators have a doubly clamped geometry, some singly clamped ones have been studied as well.[34,54] Most of them simply involve NWs or CNTs grown vertically from a substrate, though some have been assembled into more deliberate structures and explored to function as nanotube radio or mass sensors.[55,56] For the rest of this review, we will mainly focus on doubly clamped devices when discussing 1D NEMS resonators, as these devices are more relevant to device applications.

### 2.2.2. Fabrication Techniques for 2D NEMS Resonators.
2D resonators can also be fabricated in a number of different ways. One of the early approaches is similar to fabricating CNT resonators, starting with a flake on a substrate and followed by metal deposition and sacrificial layer etching, resulting in a suspended 2D ribbon as a doubly clamped resonator.[57] Some variations to this approach include using polymer such as SU-8 to build a support structure around the 2D flake (Figure 2.1c), thus forming a fully clamped resonator instead,[58,59] and using chemical vapor deposition (CVD) grown 2D materials instead

of exfoliated flakes, forming large-scale arrays of 2D NEMS resonators by prepatterning the 2D layer.[60]

Alternatively, one can use prefabricated substrates with holes/trenches and electrodes and then directly exfoliate the 2D material onto the substrate, and there is a chance that 2D flakes will be suspended over the holes/trenches. Both graphene and $MoS_2$ resonators were first achieved using this approach.[16,17] However, such approaches have some degree of randomness, and the fabrication outcome is often hit or miss.

A more deterministic alternative has been developed based on a transfer technique, in which a 2D material is first exfoliated onto a polymer stamp such as polydimethylsiloxane (PDMS), with a target flake identified.[61] Then the target flake is aligned to features on a prepatterned substrate with microtrenches and electrodes, and dry-transferred onto the targeted location on the substrate (Figure 2.1d), resulting in suspended resonator structures. Interestingly, such transfer techniuqe can also result in large arrays of 2D resonators when combined with water-assisted transfer.[62]

## 3. EXCITATION AND DETECTION OF VIBRATORY MOTION IN NEMS RESONATORS

Beyond device fabrication, another major challenge in NEMS resonator research comes from the precise measurement of the mechanical motion in these minuscule structures. NEMS resonators are typically of truly atomic scale in the direction of motion, and the motional amplitude is often on the order of nm or even smaller. It is often ineffective to simply attempt well-known motion detection schemes used for much larger, mainstream MEMS devices. Therefore, tailored excitation and measurement techniques are necessary to detect the motion of these nanoscale mechanical devices and to study their properties.

### 3.1. Excitation of Resonant Motion.
In NEMS resonator research, excitation of vibrational modes is typically achieved optically, magnetically, acoustically, or electrically.

Optical (optothermal-mechanical) excitation often uses an intensity-modulated laser to periodically heat up the material through optical absorption, thus photothermally driving the motion of the resonators through thermal expansion.[63] This technique poses minimal requirement on the device structure (e.g., not requiring electrodes) and material properties (e.g., no need to be conductive), and is suitable for fast prototyping of NEMS resonators made from emerging nanomaterials (Figure 3.1a).

Magnetomotive excitation is based on the magnetic force induced by the current flowing through the suspended portion of the device perpendicular to a magnetic field, which has been used to drive mechanical motion in Si NWs.[10] Excitation can also be achieved by integrating a piezoelectric material with the substrate, which can acoustically shake the device with the application of an AC voltage using the inverse piezoelectric effect (Figure 3.1b).[64] Some NEMS structures with one of the dimensions extended into the macroscopic scale (such as ultralong suspended CNT) can even be acoustically excited using soundwaves.[44]

Electrical excitation is probably among the most widely used techniques, which can usually be achieved in a capacitive scheme (Figure 3.1c−e).[65] By applying driving voltages between the gate and the low-dimensional material that is suspended above the gate and separated by a small gap (typically sub-$\mu$m scale), the electrostatic force can drive the suspended nanomaterial into periodic motion.[57] This usually requires the low-dimensional material to be metallic or semiconducting. For insulating materials, such as h-BN, a modified electrical excitation technique can be used: Given that the material part of the resonator is effectively a movable dielectric layer between two electrodes, one can utilize the electrostatic force based on the dielectric effect to excite the motion of the insulating material.[66]

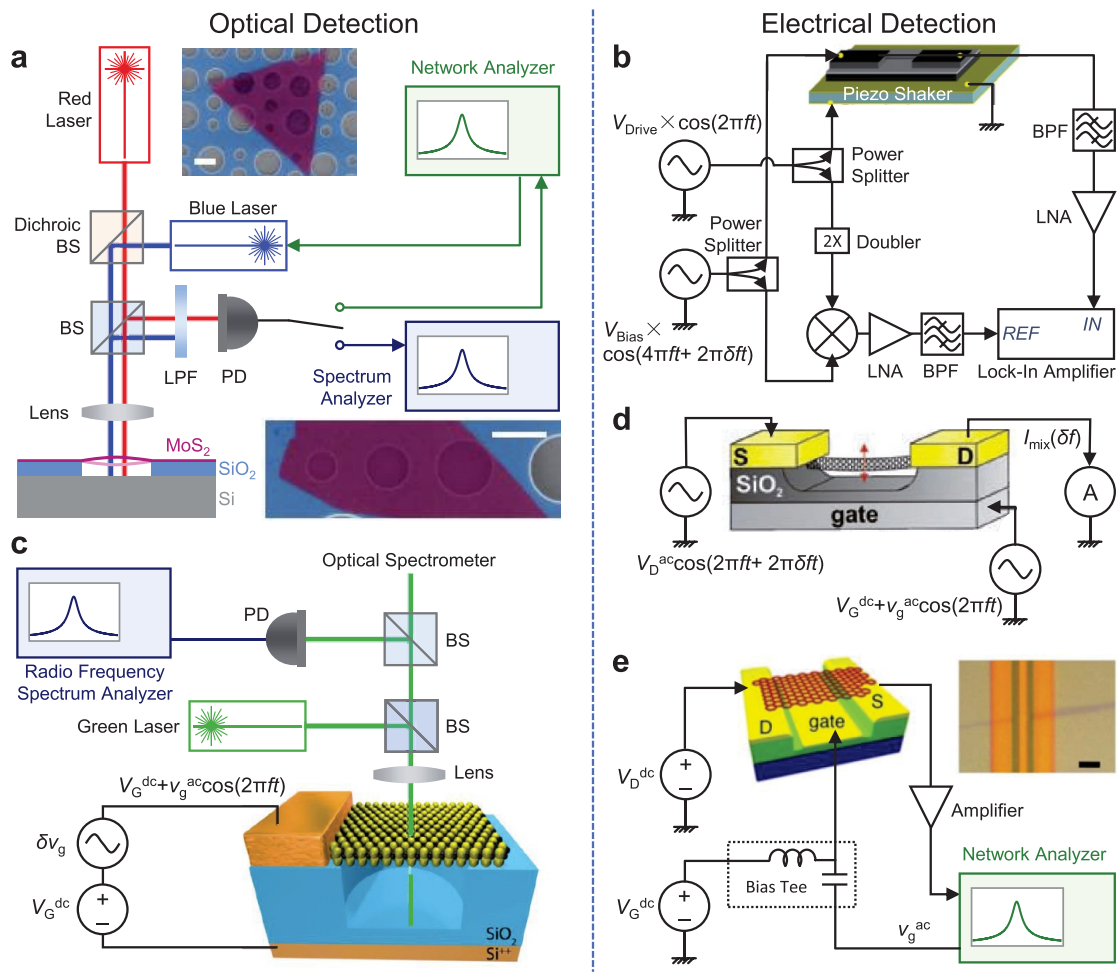

**Figure 3.1.** Nanomechanical resonance excitation and measurement techniques. (a) Schematic for optothermal excitation (blue laser) and optical interferometry readout (red laser) of 2D MoS₂ resonators. Reprinted in part with permission from ref 63. Copyright 2018 The Authors, some rights reserved; exclusive licensee AAAS. Distributed under a CC BY-NC 4.0 license http://creativecommons.org/licenses/by-nc/4.0/. (b) Piezoelectric excitation and piezoresistive detection of Si NW resonators. Reprinted in part with permission from ref 72. Copyright 2008 American Chemical Society. (c) Schematic for capacitive excitation and optical interferometry readout of 2D tungsten disulfide (WSe₂) resonators. Reprinted in part with permission under a Creative Commons (CC BY) License from ref 65. Copyright 2016 American Chemical Society. (d) Capacitive excitation and frequency down-mixing electrical readout of CNT resonators. Reprinted in part with permission from ref 110. Copyright 2009 American Association for the Advancement of Science. (e) Direct RF electrical readout of graphene resonators. Note the difference between the global gate design and the local gate design (as in d and e, respectively). Reprinted in part with permission from ref 87. Copyright 2010 American Institute of Physics. Panels (a) and (c) belong to optical detection, while panels (b), (d), and (e) belong to electrical detection.

## 3.2. Detection of Resonant Motion.

Compared with device fabrication and motion excitation, resonance detection is arguably the most technically challenging aspect in the experimental studies of NEMS resonators. Due to the much smaller device size and smaller resonance signal amplitude, existing resonance detection techniques used for MEMS resonators, such as capacitive readout, high-speed camera imaging, laser Doppler vibrometry, etc., may not be straightforwardly applicable or effective, and different (or at least carefully customized) techniques have to be developed. To date, most NEMS resonators are measured either optically or electrically.

Among optical detection schemes, laser interferometry is arguably the most widely used technique, especially in the study of 2D NEMS resonators (Figure 3.1a,c). It is based on the interference resulting from multireflection of light within the device structure: While a laser beam is incident on the resonator, the spacing between the suspended material and the substrate changes as the 2D material vibrates, and thus changes the interferometry condition of the entire device structure. The reflected light intensity is therefore modulated by the motion of the resonators, and the depth of modulation reflects the amplitude of the motion.[67] The light signal is then converted to an electrical signal using a photodetector, which contains the mechanical vibration signal. At the resonance frequency, the motion amplitude peaks, and so does the variation of the optical signal. This technique is very sensitive to device motion, with demonstrated fm/Hz^{1/2}-level sensitivity at room temperature,[68] and thus capable of measuring the undriven thermomechanical resonance induced by the Brownian motion of the nanodevice (Figure 3.1a, switch position downward, blue laser off).[16,63]

Electrical resonance detection schemes, on the other hand, are more compatible for integration with electronic circuits. However, because of the minimal device size in resonant NEMS resonators, the motional signal from the vibrational structure can often be overwhelmed by the parasitic background resulting from the much larger electrodes (often orders-of-magnitude larger than the size of the motional structure, and thus with a much larger capacitance). Therefore, some readout techniques very widely used in MEMS studies, such as direct capacitive signal transduction without mixing, hardly work for NEMS resonators. To address this challenge, frequency mixing is used to detect the resonance,[69] where a frequency-modulated (FM) or amplitude-modulated (AM) radio frequency (RF) input is often used to create signals with slightly different frequencies $\Delta f$ from the resonance frequency (Figure 3.1b,d). The resonator works as a mixer (multiplying signals of different frequencies) with its conductance modulated by its vibration, resulting in a signal at the intermediate frequency (IF, which equals $\Delta f$ in this case) that carries information on device motion.[11,70] This low-frequency signal is then amplified and measured by a lock-in amplifier, which separates it from the parasitic driving signal at high frequency. It is worth mentioning that in order for the resonator to behave as a mixer, its conductance needs to be modulated by the displacement, via changes in carrier density, piezoresistive effect, etc. Therefore, it is not uncommon for researchers to operate NEMS resonators at the transconductance peaks, such as on the edges of the Coulomb blockade peaks.[71]

In addition to optical and electrical readouts, a number of detection techniques have also been demonstrated. For example, the magnetomotive detection technique has been used for NW resonators, which monitors the electromotive force voltage as the device vibrates in a magnetic field.[10] Employing the piezoresistivity of Si, Si NW resonators have been measured using piezoresistive readout techniques.[72] Researchers have also demonstrated mechanical detection of graphene/CNT resonator motion using atomic force microscopy (AFM),[73,74] scanning and transmission electron microscopy (SEM and TEM)[75,76] or by coupling NEMS resonators to superconducting cavities.[77−79] Some of these techniques are less common for engineers as they require additional instrumentation (such as an external magnetic field or AFM/TEM/SEM) and are challenging for integration toward device applications.

**3.3. Implications from Device Structure and Dimensionality.** It is noted that most 1D resonators are measured with electrical readout, while optical interferometry technique is far more common in 2D resonator studies (see Table 1 in Section 5.6). The main reasons are as follows: Optical measurements for 1D resonator are very challenging, especially for CNT resonators. Consider a laser spot focused by an objective lens, which is typically on the order of the wavelength of the laser, i.e., a few hundred nanometers. The diameter of a CNT is orders of magnitude smaller than the laser spot, and thus only a very small fraction of reflected light carries information on the CNT motion. This significantly affects the signal-to-background ratio and the measurement efficiency, unless a scattering center is functionalized on the nanotube to enhance light reflection[44,80] or a nanotube bundle is positioned in an optical cavity.[81] In contrast, in 2D resonators, the entire laser spot can be positioned within the vibrating structure (often 1 $\mu m$ or more in lateral size), and thus can utilize the strong coupling between motion and optical field in transducing motional signal. It is worth noting that for certain NW resonators, while featuring a 1D geometry, the relatively larger diameter of the NWs (compared with CNTs) facilitates optical detection of their resonant motion.[82,83]

Another favorable factor for interferometric detection can be found in the commonly used designs of 2D NEMS resonators, which often have a small vacuum gap (usually a few hundred nm) underneath the 2D flake. The size of the gap, together with the partial transparency of 2D materials, allows sufficient light to penetrate the 2D material and then reflect from the substrate, which interferes with the light reflected from the top surface of the 2D material. This obviates the use of another mirror (such as in the reference arm in a Michelson interferometer),[84] allowing interferometry to be realized more readily in such device structures.[67] More interestingly, such a vacuum gap can even be leveraged to tune the optical detection efficiency, from optimized value all the way to zero responsivity.[85] Alternatively, for monolayers and multilayers with low reflection coefficients, a standing wave is formed by the interfering incident and reflected laser beams with the reflection from the metal gate. The monolayer is a mobile absorber in this optical standing wave, and its motion is measured by recording the reflected light intensity.[86]

While most 2D resonators are convenient for measurements using optical detection techniques, 1D resonators are highly suitable for electrical readout schemes. This is because doubly clamped device structures prevail among 1D resonators, in which *all* the current has to pass through the motional channel. As of 2D resonators, this is only true for doubly clamped devices (which are indeed mostly measured with electrical readout); for fully clamped devices, electrical current can flow between electrodes through both the suspended and the nonsuspended regions of the 2D film. Therefore, only *a portion* of the current carries information on the device vibration, which limits the weight of the motional signal in the measured current and thus the detection efficiency.

As previously discussed, another important challenge in electrical readout is the parasitic signal. While mixing is one way to mitigate such effects, it could sometimes be undesirable for signal processing as it essentially converts the resonant signal to an entirely different frequency. One way to address this is by minimizing the parasitic capacitance using a local gate structure, with which direct electrical readout of the resonance has been achieved by measuring the transmission parameter $|S_{21}|$ (Figure 3.1e).[87] In such cases, the resonator acts as a resonant channel transistor[88] and allows the feedback and amplification of the signal to be used to construct a self-sustained oscillator.[59]

# 4. DEVICE CHARACTERISTICS OF 1D RESONATORS

**4.1. Carbon Nanotube Resonators.** Among low-dimensional nanomechanical resonators, doubly clamped CNTs distinguish themselves by their ultralow mass and excellent quantum transport properties. These attributes give rise to a wealth of phenomena that are much harder to study using other devices and offer the possibility for a number of applications. While double-walled and multiwalled CNT resonators have also been realized, the above statement is particularly true for single-walled CNTs (SWCNTs). We therefore discuss singled-walled CNT NEMS resonators with greater detail in this section.

*4.1.1. Carbon Nanotubes from an Atomic Perspective.* CNTs represent the ultimate size limit of 1D mechanical resonators. Single-walled CNTs are hollow cylinders whose walls consist of a single layer of graphene, i.e., a honeycomb lattice of carbon atoms,[89] as illustrated in Figure 4.1a.

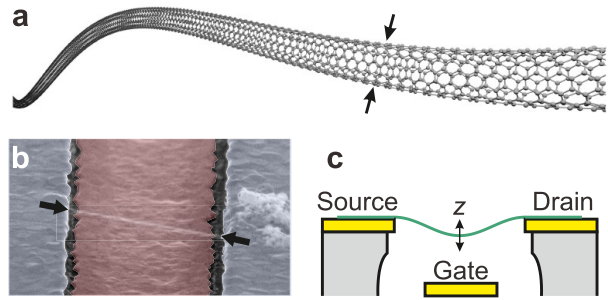

**Figure 4.1.** Structure of CNTs and typical device geometry of a nanotube resonator. (a) Illustration of the atomic lattice forming a nanotube. Individual carbon atoms are shown as spheres and carbon−carbon bonds as lines. The distance between the arrows is the tube diameter. (b) Colored scanning electron micrograph showing a nanotube suspended over a trench. The gate electrode underneath the nanotube is shown in red. The arrows indicate the clamping points. Adapted with permission from ref 98. Copyright 2011 American Chemical Society. (c) Sketch of typical device geometry, with a nanotube (green) suspended freely between source and drain electrodes over a gate electrode and vibrating in $z$-direction. Metal electrodes are shown in yellow.

previously discussed, a great majority of CNT resonators have a doubly clamped geometry (Figure 4.1b,c). When a suspended nanotube is clamped at the two ends, it vibrates like a guitar string.[11]

Most of the special properties of nanotube resonators can be understood from their special structure. (i) The mass of a nanotube is very small because it consists of only a small number of atoms. For a typical CNT with a radius of about 1.5 nm and a length of 1 $\mu$m, the mass is roughly $m = 7 \times 10^{-21}$ kg, corresponding to $3.5 \times 10^5$ carbon atoms. The fact that the cross section of a CNT consists of only few (<10 for the thinnest ones)[90] atoms is also responsible for its extremely low spring constant (for transverse deflection), down to ∼10 $\mu$N/m in the doubly clamped configuration.[91] This makes it possible to significantly tune the resonance frequency and other mechanical properties of the nanotube via external forces. (ii) The regular 2D crystal of carbon atoms rolled up into a nanotube can act as an electrical conductor with a range of fascinating features. In particular, at low temperatures the CNT can turn into a Fabry−Pérot resonator[45] or a quantum dot,[92] depending on the electron transmission between the nanotube and the electron reservoirs.[93] The coupling between such highly tunable electrical states and the nanotube motion is the basis for some of the most exciting ideas for future developments.

### 4.1.2. Mechanical Properties of CNT Resonators.
From a purely mechanical point of view, a CNT can be well approximated as a hollow beam and modeled with the corresponding Euler−Bernoulli equation.[94] The validity of this continuum model is surprising, considering that the radius of a nanotube is only about 10 times larger than the separation between atoms in the lattice. In principle, the Euler−Bernoulli model can help to identify the eigenfrequencies of discrete vibration modes. For a particular mode, the transverse vibration amplitude $z$ can be readily expressed in an equation of motion:

$$\ddot{z} + \gamma\dot{z} + \omega_0^2 z + \alpha z^3 + \beta z^2 + \eta z^2 \dot{z} = F(t)/m \quad (4.1)$$

where dots denote time derivatives, $\omega_0$ is the angular resonance frequency, $\gamma = \omega_0/Q$ describes the energy dissipation with $Q$ being the quality factor, and $F(t)$ is the external driving force. When the nanotube displacement is large enough, nonlinear spring and nonlinear damping forces become significant. In eq 4.1, $\alpha$ is the nonlinear spring constant for the cubic term of displacement (quartic and symmetrical contribution to the resonator potential energy), $\beta$ is the nonlinear spring constant for the quadratic term of displacement (cubic and antisymmetrical contribution to the resonator potential energy),[94−97] and $\eta$ is the coefficient of a nonlinear damping process.[98−101] We will discuss about nonlinearities in NEMS resonators with greater detail in Section 9.1.

### 4.1.3. Unique Features of CNT Resonators.
While a nanotube resonator can be described by the same equation of motion as the majority of NEMS resonators, it possesses a number of unique features. The resonance frequency $f_{res}$ of flexural modes in CNT resonators can be pushed to impressive high values above 10 GHz,[102,103] which is yet to be achieved in flexural-mode 2D resonators. CNT resonators can also feature low intrinsic dissipation. This is a consequence of the high crystallinity of nanotubes and their lack of dangling bonds at the surface, which largely prevents chemisorption of molecules that could act as two-level systems. When all potential sources of electrical and mechanical noise are minimized, the quality factor of ultraclean CNTs can reach 6.8 million at 70 mK.[48,104]

The demonstration of the large $Q$ initially came as a surprise; for many years, researchers have observed that quality factors would decrease with the volume of the resonator, and did not expect nanotubes to exhibit such giant quality factors, especially as no special scheme or measure for "dissipation dilution" is at work.[105−108] Therefore, it is surprising and encouraging for researchers to find that among the best-reported nanotube devices, the corresponding dissipation coefficient is on the order of $m\gamma \approx 3 \times 10^{-19}$ kg·Hz.[104] This dissipation coefficient is a crucial number for force sensing applications because it directly determines the thermal force noise power spectral density $S_F \propto m\gamma$.

Another important difference between a nanotube resonator and a typical beam is related to static deformation. In MEMS structures, the shape of a beam made from metal or dielectric is usually defined in a top-down lithographic patterning process, and its static profile cannot be easily modified in a continuous and substantial way by electrostatic forces. In contrast, the amount of slack and mechanical tension in a nanotube is difficult to control in the fabrication process and can vary strongly between nominally identical devices.[97,109] Nevertheless, a gate electrode in the microtrench under the nanotube offers a convenient tool to control the resonator tension via the electrostatic force.[11,110,94,96−98,111] This force changes the nanotube shape significantly and thus induces tensile strain in nanotube, which enables frequency tuning[11,94,96−98,109−112] (Section 7.1) and mode coupling (Section 10.1). In addition, a static force applied on a straight nanotube breaks the symmetry of a harmonic potential due to bending of the nanotube, which leads to a striking behavior where the equilibrium position of the mode can be controlled by its vibration amplitude,[97] as illustrated in Figure 4.2.

Furthermore, nanotubes exhibit excellent sensitivity to small forces. Eq 4.1 reveals that the low mass increases the acceleration of a resonator for a given force, $\ddot{z} \sim F/m$. On resonance, this relationship simplifies to a steady-state response of $z = \frac{FQ}{m\omega_0^2}$.

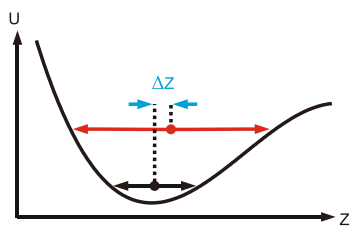

**Figure 4.2.** Sketch of an oscillation potential $U(z)$ with a broken symmetry. Small and large oscillations have different equilibrium positions, see black and red lines. A change in the oscillation amplitude results in a static shift $\Delta z$. Reprinted in part with permission from ref 97. Copyright 2013 Springer Nature.

Because of its tiny mass and stiffness, the nanotube oscillation amplitude in response to a small force is orders of magnitude larger than that of most other resonators, turning it into an excellent force sensor.[91]

**4.2. Nanowire Resonators.** Compared with nanotube resonators, NW resonators exhibit greater diversity in terms of choice of materials, fabrication methods, and thus device dimensions. Consequently, device behavior and performance also vary broadly. Figure 4.3 showcases a number of CNT and NW resonators and summarizes their basic device specifications and performance (such as frequency, quality factor, and surface-to-volume ratio). It can be seen that compared with nanotubes, NWs typically have smaller surface-to-volume ratios due to their larger cross sections, and quality factors of NW resonators are less scattered (mostly within the $10^3 - 10^5$ range). Also, quality factors tend to improve at cryogenic temperatures, which is true for both CNT and NW resonators.

*4.2.1. Device Dimensions.* Unlike CNTs, which are always chemically synthesized and grown, NWs of different materials can be fabricated using a number of methods, from top-down lithography processes to bottom-up chemical synthesis. Therefore, the lateral dimensions of NW resonators can vary a lot. For example, a top-down fabricated Si nanobeam resonator has a cross section of 300 nm × 800 nm,[32] and a Pt NW resonator made by electrodeposition of Pt on a dissolvable nanoscaffold has a diameter of 43 nm.[33] Epitaxially grown Si NWs, on the other hand, can have diameters between 60 and 100 nm.[10,35]

Arrays of NW mechanical devices have also been produced. One noticeably clever technique is to translate the ultrasmall and precise thickness control of superlattice growth into defining patterns for creating high-density planar NW arrays,[113] enabling both metallic and semiconductor NW mechanical structures after etching the substrate. Another approach is to combine top-down lithography with assembly of bottom-up grown NWs. For example, by using a bottom-up assembly method which involves electric field force effect, capillary force effect, and a liftoff process, large-area Si and Rh NW electromechanical resonator arrays over centimeter-scale chip area have been fabricated.[34]

*4.2.2. Parametric Excitation and Oscillator.* In studies of nanomechanical resonators, it is sometimes possible to achieve a reduction in the apparent resonant peak width using certain experimental schemes, with resulting phenomena termed enhancement in the *effective Q*. Such techniques have been explored for NW (and CNT) resonators. <u>Parametric pumping</u>: Parametric amplification (exciting the resonator at twice of its resonance frequency) can lead to a resonance line width narrowing, which has been demonstrated in a CNT resonator[98] and a silicon carbide (SiC) NW resonator.[114] <u>Oscillator</u>: The effective quality factor can also be improved by converting an open-loop resonator to a closed-loop self-sustained oscillator, which can further concentrate the vibrational energy in the frequency domain. This has been demonstrated in a doubly clamped SiC NW NEMS oscillator, showing resonance line width narrowing of 19 times and leading to a $Q_{osc,eff}$ of 47,580.[115] Other techniques such as reduction of the phase (or frequency) noise and strain engineering also belong to this category and can lead to effective $Q$ enhancement. We shall emphasize that such phenomena do not mean reduction in the intrinsic energy dissipation of the resonators (Section 8), and therefore may not straightforwardly lead to improvement in sensing and quantum experiments.

## 5. DEVICE CHARACTERISTICS OF 2D RESONATORS

Since the discovery of graphene, extensive research has been conducted to study the various properties of 2D materials, and 2D NEMS resonators have been under intensive study. The material choice for building 2D NEMS has soon grown beyond just graphene, with 2D semiconductors, insulators, magnetic materials, and 2D HSs joining the list one after another, offering intriguing opportunities for studying electromechanical coupling and enabling different types of NEMS-based sensors. These atomically thin NEMS resonators show great promise for realizing high-frequency, high-tunability, and ultralow-power devices.

**5.1. Graphene NEMS Resonators.** NEMS resonators based on graphene, the hallmark 2D material, have been extensively studied in recent years (Figure 5.1). Nano-mechanical analysis of the natural vibration of few-layer graphene sheets[120] dates back before the experimental realization of the first graphene mechanical resonator. The initial demonstration of a graphene nanomechanical resonator (Figure 5.1a), using optical interferometry detection, shows that the fundamental mode resonance frequencies of these doubly clamped resonators range from few MHz to ∼170 MHz with $Q$ up to ∼850, depending on the geometry of the devices.[16] Later, the first electrical readout of doubly clamped graphene NEMS resonators was demonstrated using AM mixing (Figure 5.1b), showing $Q$ up to $10^4$ at 5 K as well as frequency tuning from ∼30 to 80 MHz.[57] Due to the atomic-scale thickness, in terms of elastic behavior, monolayer and few-layer graphene resonators are typically in the membrane regime, and the resonance frequency can be efficiently tuned by in-plane tension, which can be induced by deflection and stretching caused by a DC gate voltage. Meanwhile, the intrinsic strength of graphene is up to 130 GPa, which means a very high fracture limit;[13] this corresponds to a very broad strain tuning range of the resonance frequency.[121] Arrays consisting of single-, bi- and trilayer graphene circular drumhead resonators with diameters of 1 to 1.5 μm have been made with exfoliation and all-dry transfer methods with stencil lithography electrodes, showing resonance frequency up to 136 MHz and $Q$s of ∼100 to 400 at room temperature.[122]

While the initial demonstrations on graphene NEMS resonators are mainly based on mechanically exfoliated graphene flakes, later CVD graphene has also been explored for NEMS resonators. Fully clamped circular drumhead resonators with diameters up to 30 μm (Figure 5.1e) have been demonstrated using CVD graphene film[123] and generally show higher quality factors (Section 8.2) than doubly clamped ones made of the same material under similar conditions. A sizable array of single-layer graphene NEMS resonators has been demonstrated (Figure 5.1c), including 38 CVD graphene resonators with identical geometry and showing multimode resonances.[60] CVD graphene NEMS resonators in an array can also show a narrow distribution of resonance parameters.[124]

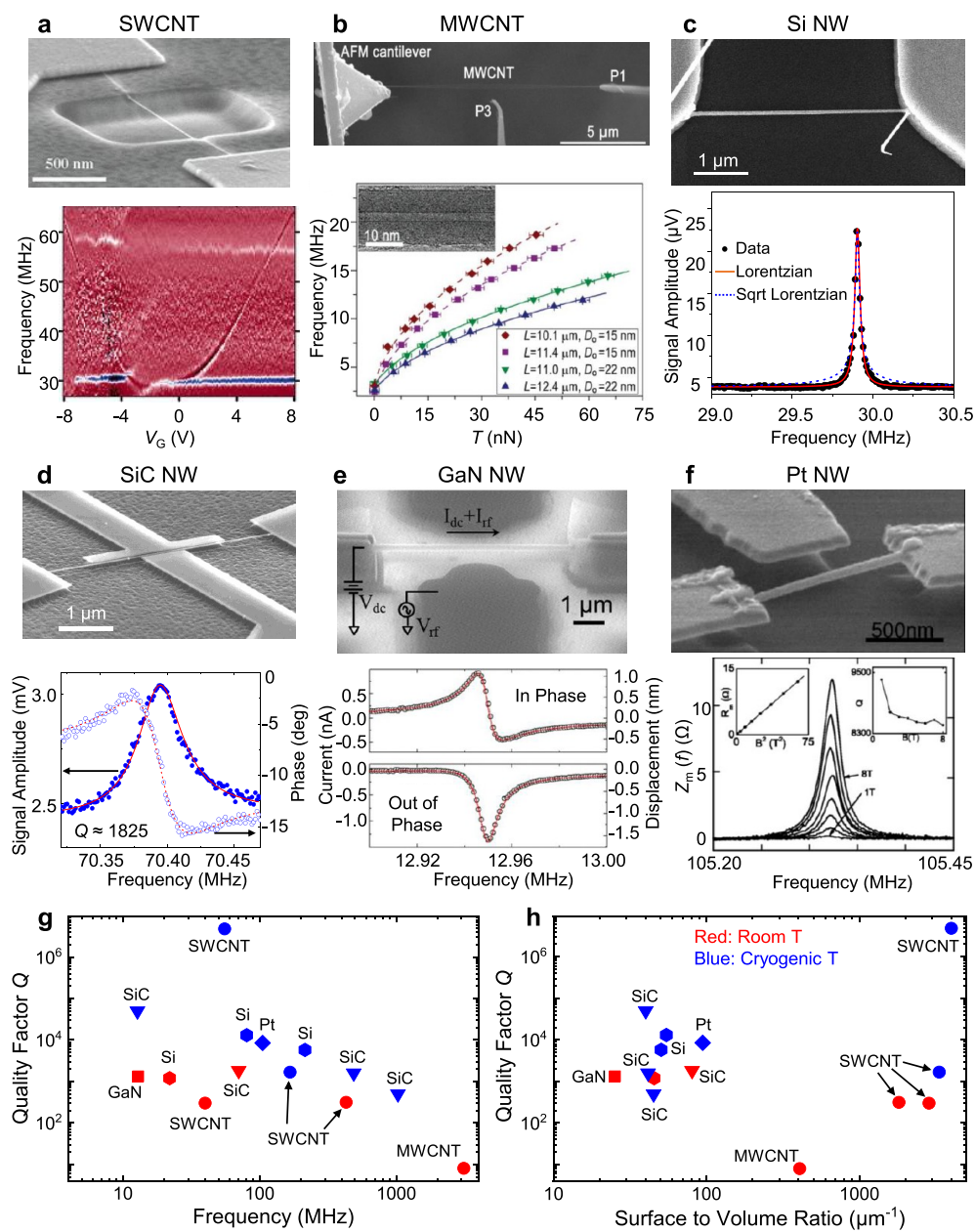

**Figure 4.3.** Examples of 1D NEMS resonators, with SEM images and representative resonance characteristics shown for (a) SWCNT, reprinted in part with permission from ref 198. Copyright 2006 American Chemical Society. (b) MWCNT, reprinted in part with permission from ref 116. Copyright 2009 Wiley-VCH Verlag. (c) Si NW, reprinted in part with permission from ref 72. Copyright 2008 American Chemical Society. (d) SiC NW, reprinted in part with permission from ref 117. Copyright 2009 Institute of Electrical and Electronics Engineers. (e) Gallium nitride (GaN) NW, reprinted in part with permission from ref 118. Copyright 2012 American Institute of Physics. (f) Pt NW NEMS resonators, reprinted in part with permission from ref 33. Copyright 2003 American Institute of Physics. (g) Measured $Q$ vs resonance frequency for some of the 1D NEMS resonators. Data taken from refs 10, 12, 33, 72, 74, 104, 117−119, 161, 198, 249, and 254. (h) Measured $Q$ vs surface-to-volume ratio for the same set of 1D NEMS resonators as in (g). In both (g) and (h), data measured at room temperature and cryogenic temperatures are shown in red and blue, respectively.

Besides resonance characteristics, graphene NEMS resonators have also been extensively used to study coupling effects and other physical processes. In one example, a graphene resonator has been capacitively coupled to superconducting microwave cavities for efficient readout of motion (Figure 5.1d). The device exhibits a $Q$ of $10^5$ at 30 mK and shows potential for studying the quantum squeezing of mechanical states.[125] Similarly, another graphene resonator coupled to a superconducting microwave cavity[78] shows a $Q$ of 220,000 at 14 mK and a displacement sensitivity of 17 fm/Hz$^{1/2}$. In another example, using Raman spectroscopy as a local optical probe, motion-induced strain and device resonance characteristics in a cantilever-shaped multi-layer graphene NEMS resonator and a circular drumhead monolayer graphene NEMS resonator have been revealed.[126,127] A GHz graphene p−n junction resonator has also been demonstrated,[128] in which the resonant signals are found to be strongly enhanced in the bipolar regime.

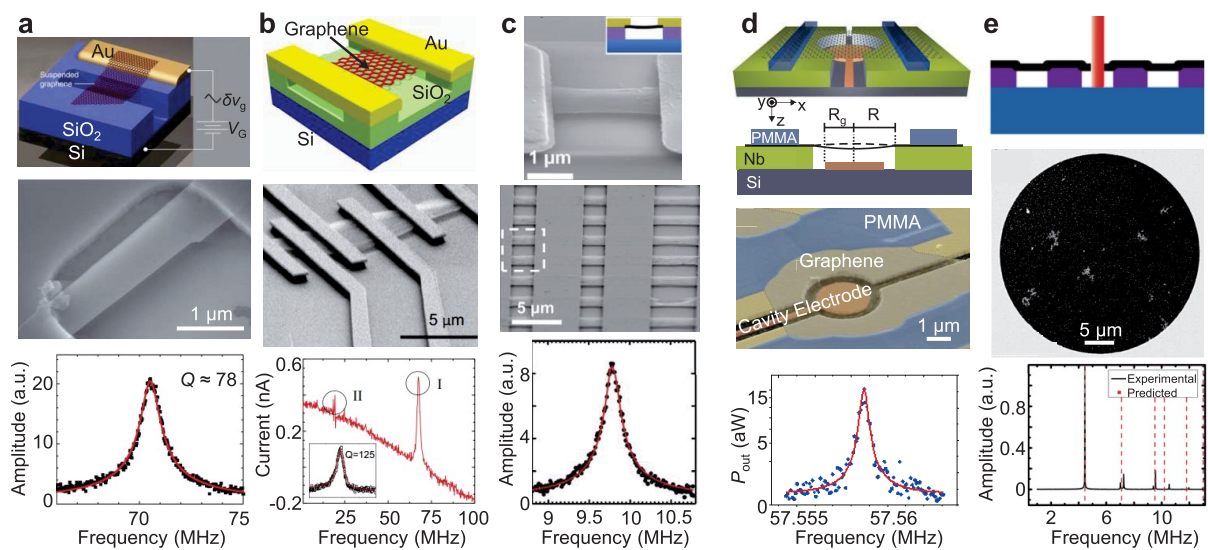

Figure 5.1. Graphene NEMS resonators with examples of device schematics, SEM images, and measured resonances, for (a) the first graphene NEMS resonator with optical interferometry readout. Reprinted in part with permission from ref 16. Copyright 2007 American Association for the Advancement of Science. (b) The first graphene NEMS resonator with electrical readout. Reprinted in part with permission from ref 57. Copyright 2009 Springer Nature. (c) A large-scale array of NEMS resonators made from CVD graphene. Reprinted in part with permission from ref 60. Copyright 2010 American Chemical Society. (d) A graphene resonator coupled to a superconducting microwave cavity. Reprinted in part with permission from ref 125. Copyright 2014 American Chemical Society. (e) CVD graphene resonators with diameter up to 30 μm showing size-dependent quality factor. Reprinted in part with permission from ref 123. Copyright 2011 American Chemical Society.

## 5.2. 2D Semiconductor NEMS Resonators.

Beyond graphene, 2D semiconductors such as TMDCs and black P also provide attractive electrical, optical, and mechanical properties, and are extensively explored for NEMS devices.[17−20,65] For example, a theoretical calculation suggests that $MoS_2$ resonators could have lower energy dissipation and thus higher Q factors than their graphene counterparts.[129] $MoS_2$ also shows a larger piezoresistive gauge factor compared with graphene, thus $MoS_2$ NEMS resonators could exhibit stronger electromechanical coupling, as its device motion not only modulates the channel conductance through the gating effect but also through the piezoresistive effect.[130]

The first $MoS_2$ nanomechanical resonator[17] (Figure 5.2a) has been measured using laser interferometry, which shows thermomechanical resonance frequencies up to 60 MHz in the very high frequency (VHF) band, a Q up to 710 at room temperature, and a frequency-Q product up to $2 \times 10^{10}$ Hz. Frequency scaling laws of $MoS_2$ are also elucidated and developed with quantitative parameters, providing important guidelines for scaling the frequency toward GHz resonators.[17] Besides thermomechanical motion, optothermally or capacitively driven resonances of 2D $MoS_2$ resonators up to 120 MHz have also been measured using optical interferometry, and by gradually increasing the driving amplitude toward mechanical nonlinearity (Section 9.1), a large linear dynamic range (DR) up to 110 dB has been experimentally demonstrated.[63] This is in clear contrast to CNT resonators, which have been expected to have very limited DR (Section 9.2).[95] Electrical readout has also been demonstrated in 2D $MoS_2$ NEMS resonators, using FM mixing and piezoresistive readout schemes.[131−133] Raman spectroscopy has been coupled with 2D $MoS_2$ NEMS resonators, which show dynamical phonon softening and Raman signal amplitude enhancement during the nonlinear resonances.[134]

Besides TMDC, other 2D semiconductors also exhibit interesting and unique properties. For example, black P

NEMS resonators have been demonstrated (Figure 5.2b) with initial devices capable of vibrating at up to 100 MHz.[19] Interestingly, due to its corrugated lattice structure, black P is intrinsically and strongly anisotropic in-plane in its electronic, elastic, optical, and thermal properties along the zigzag and armchair directions, which gives rise to resonant responses different from those made of elastically isotropic 2D materials.[135] Leveraging this unique feature of strong intrinsic anisotropy in black P and other 2D materials sharing the similar crystal structures, researchers have successfully determined its anisotropic Young's moduli by measuring the vibration frequencies and mode shapes of higher order modes (see Section 6 for more details).[136]

As in the case of 1D NEMS resonators, certain experimental techniques can be used to enhance the *effective* Q of 2D NEMS resonators, which does not necessarily suggest a reduction in the energy dissipation rate. For example, using parametric pumping, boosting of effective Q from 61 to 11 million by increasing the pump voltage has been demonstrated in an $MoS_2$ resonator.[137] Turning the open-loop resonator to closed-loop feedback oscillators also enhances the effective Q, which has been demonstrated for both graphene and $MoS_2$ oscillators.[59,266]

## 5.3. Wide-Bandgap 2D Material NEMS Resonators.

Atomic layers of h-BN crystal, the hallmark 2D dielectric, are excellent candidates as structural materials for enabling NEMS resonators, due to their outstanding mechanical properties (Young's modulus theoretically predicted to be as high as $E_Y \approx 780$ GPa and a very high breaking strain limit of $\varepsilon \approx 22\%$)[138,139] and ultrawide bandgap (5.9 eV),[140] which makes them suitable for ultraviolet (UV) signal detections. Dry-transferred doubly clamped h-BN membrane and circular drumhead resonators (Figure 5.2 c) have both been demonstrated,[20] vibrating at high and very high frequencies (from ~5 MHz to ~70 MHz). From the drumhead h-BN resonator with a diameter of ~11.3 μm and thickness of ~10 nm, thermomechanical resonances with up to 4 modes have been measured. Combining measurements and modeling of the multimode resonances, the elastic behavior of h-BN circular resonators is resolved, including the transition from the membrane regime to the disk regime, with built-in tension ranging from 0.02 N/m to 2 N/m. The Young's modulus of h-BN is determined to be $E_Y \approx 392$ GPa from the measured resonances.[20,141] It is worth noting that measured elastic moduli of 2D materials can sometimes be notably different from theoretical values (such as in the case of black P[135,136] as well), suggesting that nanomechanical resonant measurements can offer important insight into mechanical properties of 2D materials, complementing first-principle calculations.

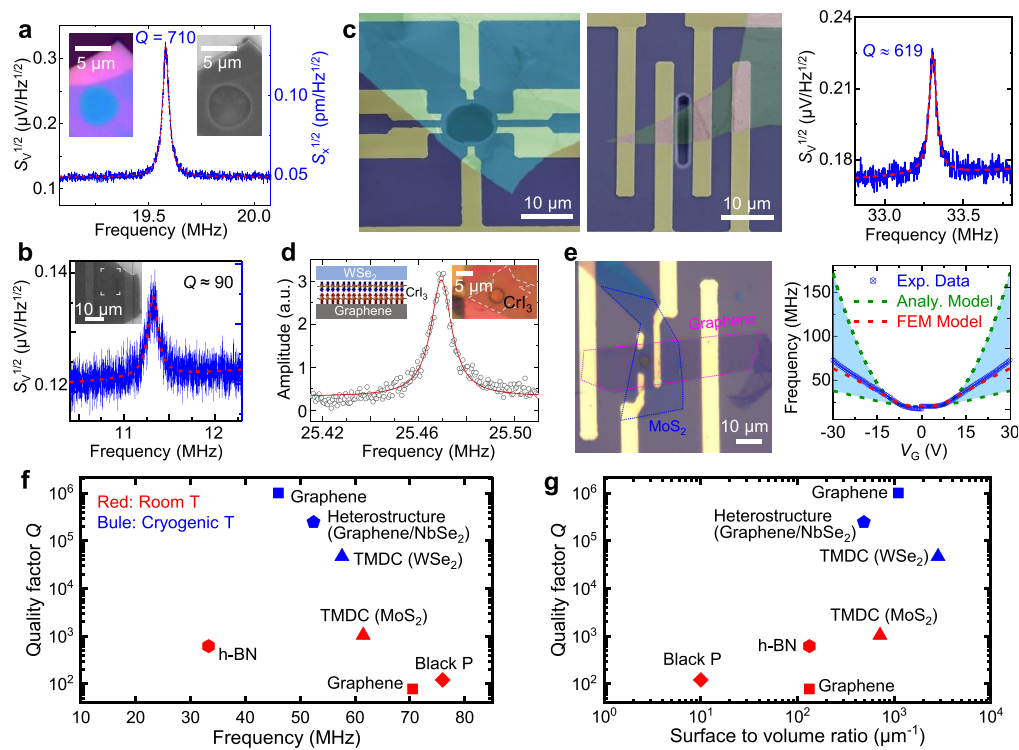

Figure 5.2. NEMS resonators based on 2D materials beyond graphene. (a) The first experimentally demonstrated MoS$_2$ nanomechanical resonator, showing undriven thermomechanical motion. Reprinted in part with permission from ref 17. Copyright 2013 American Chemical Society. (b) The first black P NEMS resonator. Reprinted in part with permission from ref 19. Copyright 2015 Royal Society of Chemistry. (c) False-colored SEM images of circularly clamped and doubly clamped h-BN resonators as well as the undriven thermomechanical resonance spectra. Reprinted in part with permission under a Creative Commons (CC BY) License from ref 20. Copyright 2017 Springer Nature. (d) Resonant response of a NEMS resonator based on 2D antiferromagnets chromium triiodide (CrI$_3$) encapsulated by WSe$_2$ and graphene. Reprinted in part with permission from ref 21. Copyright 2020 Springer Nature. (e) NEMS resonator based on 2D atomic layer van der Waals HS and gate tuning of its resonance frequency. Reprinted in part with permission from ref 179. Copyright 2021 American Chemical Society. (f) Measured $Q$ vs resonance frequency for some representative 2D NEMS resonators. Data taken from refs 16, 19, 20, 63, 142, and 154. (g) Measured $Q$ vs surface-to-volume ratio for the same set of 2D NEMS resonators as in (f). In both (f) and (g), data measured at room temperature and cryogenic temperatures are shown in red and blue, respectively. Some representative data from graphene NEMS are included in (f) and (g) for reference.

Just as semiconductors with 3D crystals (e.g. silicon) can be grown in the form of 1D NWs, wide-bandgap materials can also be grown into low-dimensional nanostructures. One such example is beta gallium oxide ($\beta$-Ga$_2$O$_3$), an emerging ultrawide-bandgap (~4.8 eV) semiconductor with applications in power and RF electronics, solar-blind UV optoelectronics, and gas sensors.[143] Its anisotropic crystal structure allows controlled growth of low-dimensional nanostructures such as NWs and nanoflakes.[144] With its excellent mechanical properties (typical Young's modulus $E_{Y,\beta\text{-Ga}_2\text{O}_3} = 261$ GPa and speed of sound $c_{\beta\text{-Ga}_2\text{O}_3} = 6623$ m/s), $\beta$-Ga$_2$O$_3$ crystal has been successfully demonstrated for NEMS resonators,[144–146] with frequency potentially beyond 4 GHz by tuning the device geometry and dimensions.[144] Electrical readout of $\beta$-Ga$_2$O$_3$ vibrating channel transistors has been demonstrated, showing potential for integration with $\beta$-Ga$_2$O$_3$ power and RF electronics.[147] Resonant transducers for solar-blind UV detection with high responsivity and fast response time have also been explored for $\beta$-Ga$_2$O$_3$ NEMS.[148,149]

### 5.4. 2D Magnetic Material NEMS Resonators.
2D magnetic materials, such as iron phosphorus trisulfide (FePS$_3$), manganese phosphorus trisulfide (MnPS$_3$), and nickel phosphorus trisulfide (NiPS$_3$), etc., have been explored to enable 2D magnetic NEMS resonators. Using their resonant characteristics, magnetic phase transitions in these materials have been studied. For example, in a study of NEMS resonator based on 2D antiferromagnetic chromium triiodide (CrI$_3$), the frequency shift is found to be related to the magnetic state of the material (Figure 5.2d). In addition, the magnetostriction effect has been quantified, and strain tuning of the magnetic interaction has been realized.[21] In another example, using temperature-dependent measurements, shifts in the resonance frequency ($f_{res}$), $Q$, and $\frac{d(f_{res}^2(T))}{dT}$ (translating to heat capacity) have been observed, revealing the transition between antiferromagnetic and paramagnetic phases in 2D FePS$_3$, MnPS$_3$ and NiPS$_3$.[22] These findings demonstrate the possibility of detecting and even controlling the magnetic states and phase transitions in 2D magnets using NEMS resonators.

**Table 1. Summary of 1D and 2D NEMS Resonators**

| | Material | Detection Techniques | Notable Values of $f_{res}$ Tuning Range $\Delta f_{res}/f_{res}$ | Notable $Q$ Values | | Linear Dynamic Range (DR) |
|---|---|---|---|---|---|---|
| | | | | Room $T$ | Low $T$ | |
| **1D** | *Carbon Nanotube* | Electrical detection[11,41,42,43,71,94,96,97,98,110] Atomic force microscopy[74] Optical cavity[217] Optical detection[159,80] | 2,000% (Mechanically deforming the substrate)[12] 330% (Electrostatic gating)[11] | 312 [ref 12] | 6.8 million (70 mK)[48] | Up to 65 dB for SWCNT (Calculation)[95] Up to 120 dB for MWCNT (Calculation)[95] |
| | *Nanowire* | Electrical detection (piezoresistive, capacitive[160]) Magnetomotive[10,32] Optical detection[82,83] | 466.7% (Mechanically bending the substrate)[199] 1.5% (Electrostatic gating)[96] | 390,000 [ref 199] | 36,200 (20 mK)[161] | 90–110 dB [ref 162] |
| **2D** | *Graphene* | Optical interferometry readout[16,60,86,123,163,228] Electrical readout (AM and FM mixing, direct RF readout, microwave cavity, *etc.*)[57,58,59,87,88,101,125,227,282] Atomic force microscopy[73] | 1,300% (Joule heating)[181] 240% (Electrostatic gating)[57] | 7,723 [ref 164] | 1 million (15 mK)[142] | 63–73 dB (200 K)[165] 40 dB (300 K)[123] |
| | *TMDC* | Optical interferometry readout[17,18,63,65,134,137,170,171,208,266,310] Electrical readout (FM, AM mixing)[131,132] | 130% (Thermal)[166] 75% (MEMS comb-drive stretching)[183] 230% (Electrostatic gating)[310] | 1,050 ($MoS_2$)[63] | 47,000 ($WSe_2$, 3.5 K)[65] | 70 dB [ref 63] |
| | *Black P* | Optical interferometry readout[19,136,168] Electrical readout[167] | 163% (Electrostatic gating)[136,168] | 120 [ref 19] | N/A | N/A |
| | *h-BN* | Optical interferometry readout[20,66,145,146] | 0.6% (MEMS comb-drive stretching)[183] | 619 [ref 20] | N/A | N/A |
| | *Heterostructure* | Optical interferometry readout[155,156,157] Electrical readout[154,158] | 370% (Graphene/$MoS_2$, Electrostatic gating)[179] | 700 (Graphene/$MoS_2$)[156] | 245,000 (Graphene/$NbSe_2$, 8.5–40 mK)[154] | N/A |

**5.5. van der Waals Heterostructure NEMS Resonators.** Stacking different 2D layered materials into van der Waals (vdW) HSs offers an additional degree of freedom to tailor the properties of the devices, which is being actively explored and exploited in 2D electronics and photonics.[150−153] In the mechanical domain, the introduction of vdW HSs can also lead to interesting device properties. For example, for some air-sensitive 2D materials, environmentally stable 2D materials (such as graphene or h-BN) can be used to encapsulate and thus protect the resulting 2D resonators, enhancing the mechanical property and stability of the devices.[154] In addition, 2D vdW HS resonators offer the opportunity to study the interlayer friction interactions and interfacial mechanics,[155,156] which can offer important insights into the dissipation in 2D NEMS devices.

The stacking of different 2D materials has led to NEMS resonators with intriguing device performance. For example, $MoS_2$/graphene vdW HS resonators exhibit fundamental mode resonance close to 100 MHz[157] and a very wide fractional frequency tuning range of $\Delta f/f$ (Figure 5.2e).[179] In another example, an h-BN/graphene HS resonator shows a frequency tuning range of about 100%.[158] By coupling graphene/niobium diselenide ($NbSe_2$)/graphene resonators to a superconducting cavity, very high $Q$ up to 245,000 at 8.5mK in mechanical resonance at >50 MHz has been reported.[154] These findings suggest that by designing different vdW HSs, it is possible to create resonant devices with tailored and enhanced mechanical, electrical, and optical properties.

**5.6. Summary of Device Metrics in 1D and 2D NEMS Resonators.** The properties of 2D NEMS resonators based on different materials are summarized in Figure 5.2f.g. Table 1 also summarizes some of the key information on 1D and 2D NEMS resonators studied to date, including the nanomaterial of choice, resonance detection technique, and some of the notable device performances reported. A few interesting trends can be observed. One of the most obvious features is that low-dimensional resonators are generally characterized by their large frequency tuning range. This is related to the outstanding stretchability of these materials and their low stiffness. For example, both 1D and 2D resonators have shown frequency tuning range over 1000%, which is unimaginable in most MEMS resonators lithographically patterned and machined out of bulk materials (such as Si) in the top-down paradigm. Another noticeable revelation is that 1D and 2D resonators can achieve large DR (e.g., up to 110 dB in $MoS_2$ 2D NEMS)[63] and also operate in the very deep nonlinear regime, which means that they can vibrate over a relatively large range of amplitude before reaching the breaking/fracturing limit. In terms of quality factor, Q values as high as 6.8 million has been demonstrated in 1D resonators, and Q up to 1 million is also demonstrated for 2D resonators, both at mK-level cryogenic temperatures. Clear differences in Q can be observed between room temperature and low temperature, suggesting that phonon may play a role in the dissipation process.

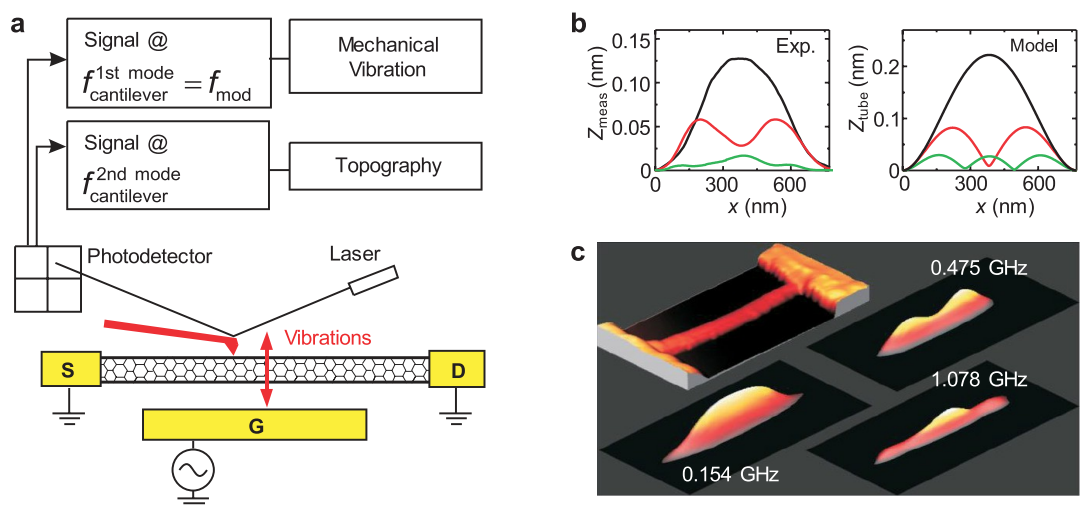

**Figure 6.1.** Visualizing motion and mode shapes in 1D NEMS resonators. (a) Schematic illustration of mode shape mapping for a CNT resonator. (b) Measured and simulated mode shapes for the first three modes. (c) 3D illustration of the measured mode shapes. Reprinted in part with permission from ref 74. Copyright 2007 American Physical Society.

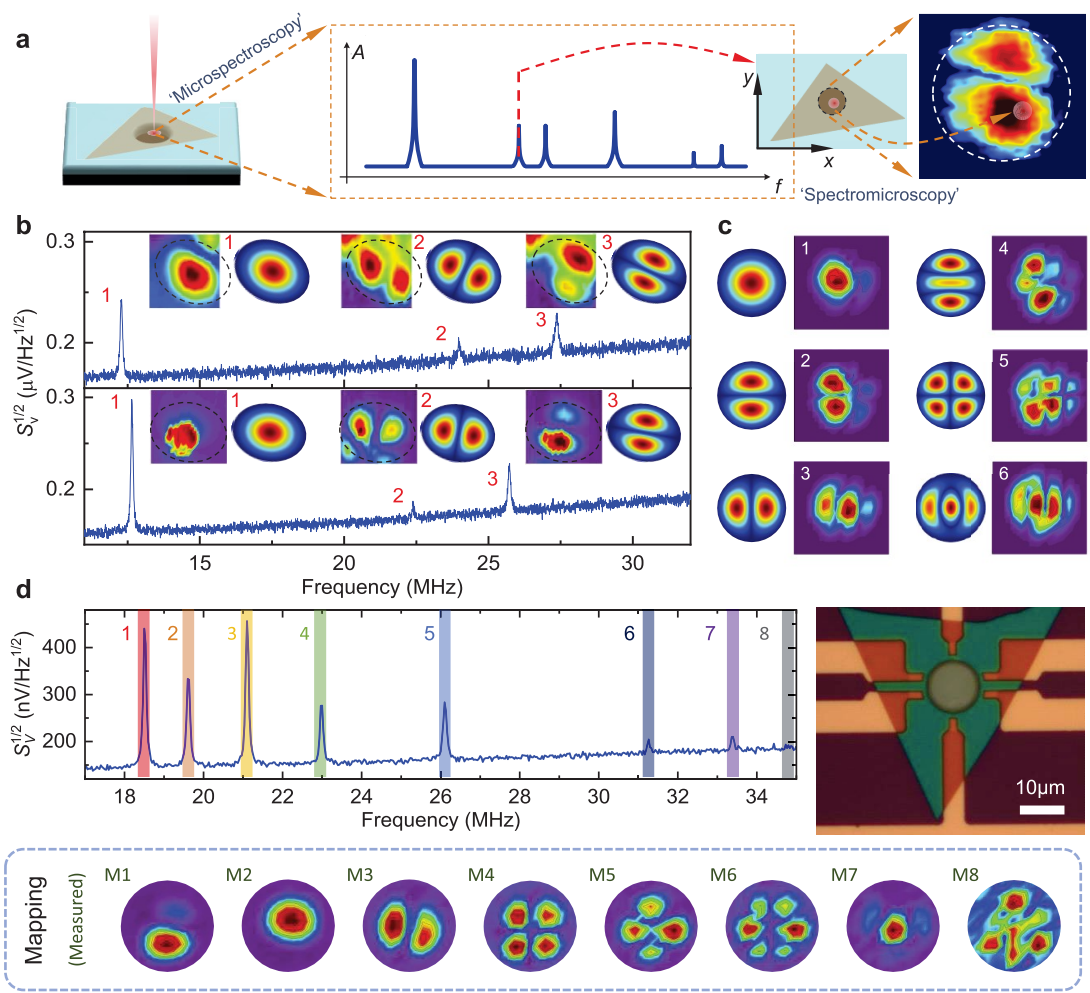

**Figure 6.2.** Visualizing motion and mode shapes in 2D NEMS resonators. (a) Schematic illustration of mode shape mapping in 2D NEMS resonators. (b) Spatial mapping of a MoS$_2$ resonator with structural nonidealities. Reprinted in part with permission from ref 170. Copyright 2014 Springer Nature. (c) Comparison of the spatially resolved mode shapes with simulation for an anisotropic black P resonator. Reprinted in part with permission from ref 136. Copyright 2016 American Chemical Society. (d) An h-BN resonator device image, measured thermomechanical resonance spectrum with 8 resonance modes, and spatial mapping of the 8 modes. Reprinted in part with permission under a Creative Commons (CC BY) License from ref 20. Copyright 2017 Springer Nature.

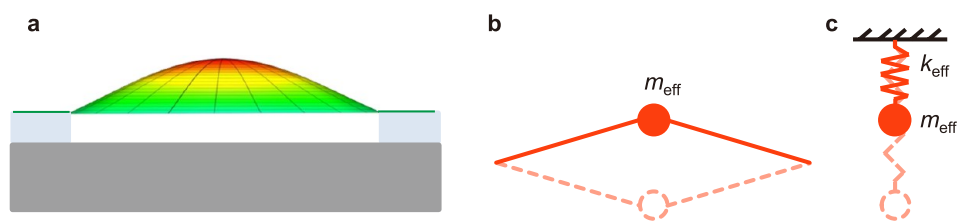

**Figure 7.1.** Mechanical model of a 2D resonator. (a) Schematic illustration (side view) of a fully clamped 2D NEMS resonator in its fundamental mode. (b) A simplified version of the resonator showing the vibration of the effective mass. (c) The lumped parameter model of the resonator in a spring-mass system.

## 6. VISUALIZING MOTION AND VIBRATIONAL MODES

It has always been highly desirable to visualize the motion of NEMS and their resonance mode shapes, and to deterministically assign and understand the multiple modes in the measured frequency-domain spectra. Specifically, the mode shape provides information about the spatial distribution of the vibrational displacement within the device, i.e., which part vibrates more and which part vibrates less, and the direction each part moves. Typically, suspended 1D or 2D NEMS resonators exhibit flexural or bending vibration, with the suspended region vibrating around the equilibrium position. Depending on the mode (except the fundamental modes), there exists one or more nodal points/lines where the displacement is 0, across which the structure moves in opposite directions at any given moment. The vibration amplitude is the largest at the antinodes. While mode shapes and mode sequence can be theoretically analyzed and numerically calculated given the boundary conditions, e.g., visualized in modeling tools such as COMSOL, the results can deviate (sometimes significantly) from the measured mode shapes and mode sequence, as geometric nonidealities and uncontrolled effects from strain affect the eigenfrequencies and the shape of the eigenmodes.

For 1D CNT resonators, mode shapes can be experimentally mapped by scanning a tapping-mode nanoprobe (i.e., in AFM) across a doubly clamped CNT while monitoring the height change (Figure 6.1). This technique has been demonstrated for a MWCNT resonator with a length of 770 nm and a resonance frequency up to 3.1 GHz.[74] Alternatively, if the device size and motion amplitude are sufficiently large, one can directly observe the resonance in a microscope. This has been done for a doubly clamped millimeter-long CNT resonator by depositing titanium dioxide ($TiO_2$) particles on the nanotube to make it visible in an optical microscope.[44] In this particular device, the resonance frequency is only a few hundred Hz. For NWs resonators and arrays with sufficiently large diameters so that the device motion can be detected optically, resonant mode shapes can also be mapped by scanning a laser spot across the entire device and measuring the vibration amplitude at each position.[169] For the less common singly clamped geometry, such as CNT cantilevers, the vibration mode shape can be resolved by measuring the intensity of inelastically scattered electrons inside an SEM.[75]

Compared with 1D NEMS resonators, 2D resonators typically have a large enough area for scanning the laser around the device and thus providing spatial resolution in resonance measurements. Therefore, spatial mapping of the mode shapes of 2D NEMS resonators is usually performed by recording the resonance signal at each location as the laser is scanned across the suspended membrane (Figure 6.2a) at each identified resonance mode frequency (i.e., "spectromicroscopy" measurement).[68] For example, spatial mapping has been performed on circular graphene resonators with diameters up to 22.5 $\mu$m, showing clearly the first few flexural modes of the circular membrane.[123] While in theory circular membranes have highly symmetric mode shapes, uncontrolled strain within the resonator, nonideal clamping at the edges, and geometrical nonidealities during device fabrication often occur, which can be effectively revealed using spatial mapping (Figure 6.2b).[170]

In addition to the device geometry, measurements of mode shapes can offer important insights into other aspects of NEMS resonators, such as resonant dynamics, material anisotropy, and built-in tension. For example, degenerate modes in $MoS_2$ resonators are distinguished to help analyze the mode coupling behavior,[171] and phase information can be derived from spatial mapping to offer insights into the complicated dynamics of graphene resonators.[163] For anisotropic 2D materials such as black P, experimental measurements of mode shapes and frequencies of higher-order resonant modes can be combined to resolve the material's anisotropic Young's moduli (Figure 6.2c).[136] Even for nominally isotropic 2D materials, such as h-BN, spatial mapping can reveal subtle structural and mechanical properties of the suspended diaphragms, including built-in anisotropic tension and bulge (Figure 6.2d),[20] thus offering guidelines on how these effects can be exploited for engineering multimode resonant functions in 2D NEMS transducers.

## 7. FREQUENCY TUNING

The capability of tuning resonance frequencies is very important for applications such as voltage-controlled oscillators, tunable RF filters, frequency-shift-based signal processing and sensing, and postfabrication reconfiguration and adjustment. The resonance frequency $f_{res}$ for MEMS/NEMS resonators (in fact, all harmonic resonators) can be generally described by the simple lumped model of a spring-mass system (Figure 7.1). In the simplest case when nonlinearities can be neglected (i.e., eq 4.1 reduces to that of a simple harmonic resonator), we obtain

$$f_{res} = \frac{1}{2\pi} \sqrt{\frac{k_{eff}}{m_{eff}}} \tag{7.1}$$

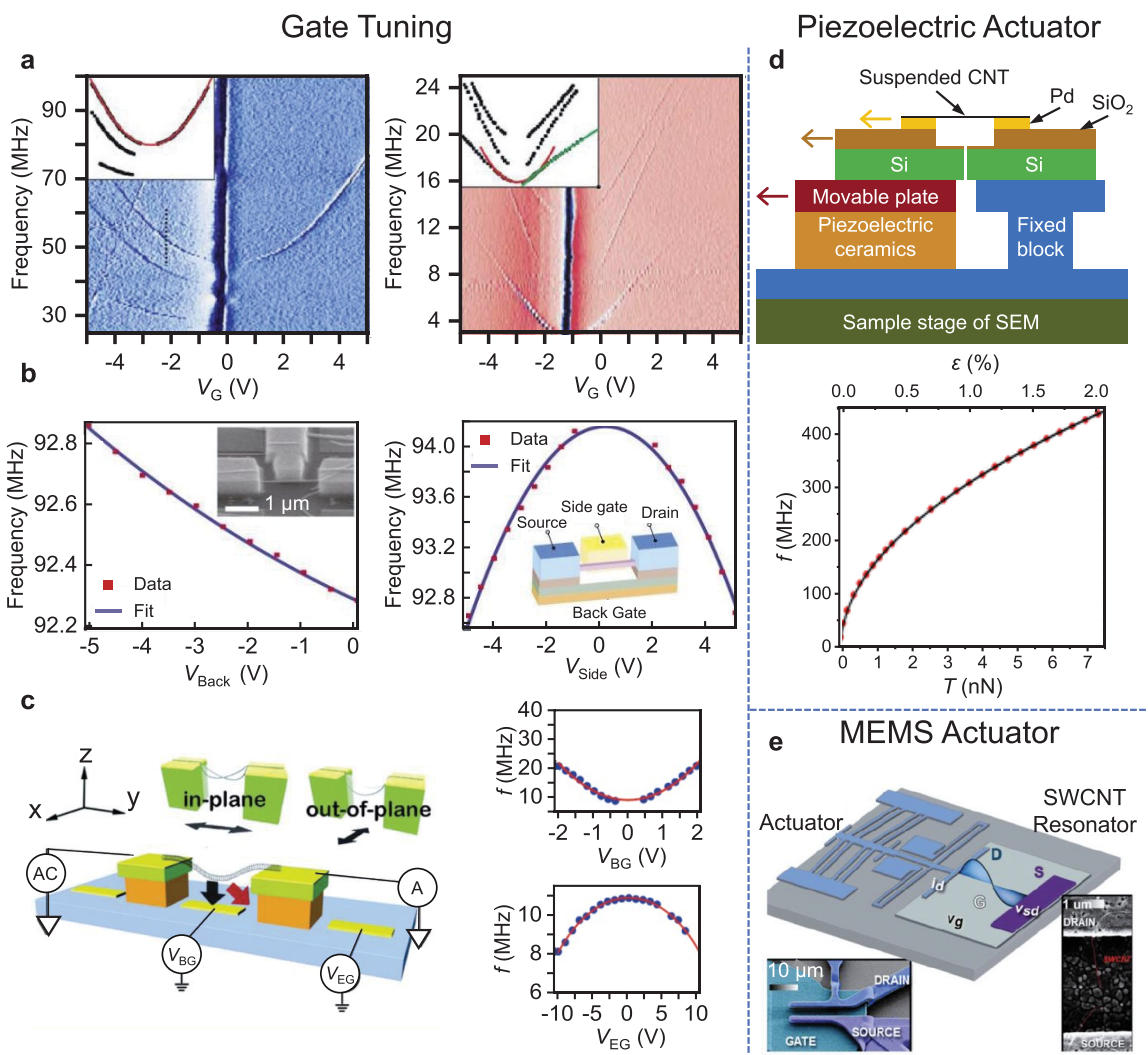

**Figure 7.2.** Frequency tuning in 1D NEMS resonators. (a) Observation of gate tuning of the resonance frequency in CNT resonators. Reprinted in part with permission from ref 11. Copyright 2010 Nature Publishing Group. (b) Capacitive softening and spring hardening in an SnO₂ NW, by sweeping the voltage on the back gate (left) or the side gate (right). Reprinted in part with permission from ref 36. Copyright 2009 American Institute of Physics Publishing. (c) Capacitive softening and spring hardening in a CNT resonator, by choosing the out-of-plane mode (top) or the in-plane mode (bottom). Reprinted in part with permission from ref 173. Copyright 2011 American Chemical Society. (d) Frequency tuning of CNT resonators using a piezoelectric actuator. Reprinted in part with permission from ref 12. Copyright 2014 American Chemical Society. (e) Axially tunable CNT resonators using cointegrated microactuators. Reprinted in part with permission from ref 111. Copyright 2014 American Chemical Society.

where $k_{eff}$ and $m_{eff}$ are the effective spring constant and effective mass of the device, respectively. Low-dimensional resonators can have large frequency tuning ranges because the $k_{eff}$ can be efficiently modulated by applying gate voltages, strain, heat, etc.

**7.1. Frequency Tuning in 1D NEMS Resonators.** Gate-induced frequency tuning has been extensively observed in 1D NEMS resonators (Figure 7.2). As previously discussed, due to their extreme aspect ratio, doubly clamped CNT resonators often contain strain built in during fabrication and are thus treated mechanically as strings (Figure 7.2a), while thicker NWs behave like beams (Figure 7.2b). Generally, strings are more susceptible to tension (in terms of resonance frequency) and thus typically exhibit larger tuning ranges than beams (compare the numbers in Figure 7.2a,b), meanwhile, slack in a string can lead to a more complicated behavior than in a beam. Specifically, doubly clamped CNTs are often not straight due to their ultralow stiffness, so that the nanotube portions near the two

clamps are not parallel during fabrication. This complicates the behavior of the frequency tuning.[94]

For simplicity, here we use a beam as an example to illustrate the frequency tuning effect from the gate. In such cases, the effect of a DC gate voltage $V_G$ on the resonance frequency can be expressed as[172]

$$f_{res,1D} = \frac{1}{2\pi}\sqrt{\underbrace{\left(\frac{E_Y I}{3\rho S} + \frac{E_Y z_s^2}{16\rho}\right)}_{\omega_0}\underbrace{\left(\frac{2\pi}{L}\right)^4}_{hardening} + \underbrace{\frac{T_0}{3\rho S}\left(\frac{2\pi}{L}\right)^2}_{\omega_0} - \underbrace{\frac{C_2 V_G^2}{\rho S}}_{softening}}$$

(7.2)

where $E_Y$ is Young's modulus, $I$ is the moment of inertia about the longitudinal axis of the beam, $L$ is the length, $S$ is the cross-sectional area of the beam, $\rho$ is the mass density of the suspended material, $T_0$ is the initial built-in tension of the as-fabricated beam (when it is not deflected due to a gate voltage), $V_G$ is the DC gate voltage, $z_s$ is the static deflection at the center point of the beam (which increases with $|V_G|$), and $C_2$ is the second Taylor expansion coefficient of the capacitance per unit length expanded in terms of the beam displacement.

When a DC gate voltage $V_G$ is applied, $f_{res,1D}$ may increase with the magnitude of $V_G$, showing a "U" shape, or decrease with $|V_G|$ instead, showing an inverted "U" shape, or even first decrease and then increase, showing a "W" shape (Figure 7.3).

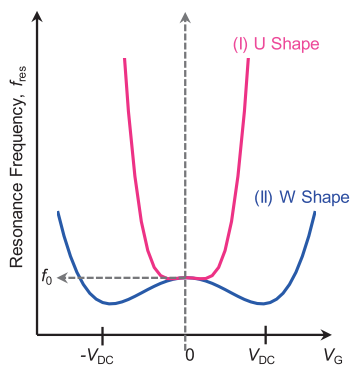

**Figure 7.3.** Schematic illustrations of tuning curves of "U-shape" and "W-shape" in 1D or 2D NEMS resonators. Reprinted in part with permission from ref 63. Copyright 2018 The Authors, some rights reserved; exclusive licensee AAAS. Distributed under a CC BY-NC 4.0 license http://creativecommons.org/licenses/by-nc/4.0/.

This is because when $V_G$ is applied, there are two effects acting on the resonance frequency. As represented by the $z_s^2$ term of eq 7.2 (electrostatic tensioning, i.e., the "hardening" term), since the voltage-induced static deflection $z_s$ will increase with $|V_G|$, it leads to a larger deflection-induced strain, which increases the resonance frequency. At the same time, the application of $V_G$ will also result in capacitive softening (the "softening" term). The competition between these two effects determines the overall shape of the frequency dependence on the applied $V_G$. While the details of the equation for a string are quantitatively different from that for a beam, the same conclusions qualitatively hold for CNT resonators, and the effects mentioned above have been observed (Figure 7.2).

These two frequency tuning mechanisms can be effectively controlled via properly designed device structures. For example, the frequency of a doubly clamped NW resonator can be tuned up or down by applying voltages to a side gate or a back gate (Figure 7.2b),[36] by leveraging the spring hardening or softening mechanisms. The side gate is in the direction of NW vibration, and thus its effect in capacitive softening is much more pronounced. Interestingly, such differentiation in frequency tuning can also be achieved by selecting different resonant modes (Figure 7.2c): those modes with device motion in the direction of the electric field lines are generally more susceptible to the softening effect.[36,173]

Besides a gate voltage, one can also control the device tension mechanically by using a piezoelectric substrate[12] (Figure 7.2d) or MEMS actuator[111] (Figure 7.2e), which can pull the resonator in-plane, achieving frequency tuning ranges up to 2000%. An interesting implication from such frequency tuning experiments is that high-frequency devices can be achieved by engineering the strain in the resonator,[11,12,111,174] in addition to scaling down the device size and choosing materials with high elastic modulus.[33,175]

## 7.2. Frequency Tuning in 2D NEMS Resonators.
For most 2D NEMS resonators made of one-to-few layer 2D materials, the membrane model typically applies, and thus the resonance frequencies are highly tunable by strain or stress.[60,176] Different from 1D NEMS resonators which are usually doubly clamped, 2D NEMS resonators can be doubly clamped or fully clamped, and the different boundary conditions can affect the frequency tuning through gating. Using a fully clamped circular membrane resonator ($f_{res,FC}$, eq 7.3) and a doubly clamped nanoribbon resonator ($f_{res,DC}$, eq 7.4) as examples, the frequency tuning by gate-induced tensile strain and capacitive softening can be written as[177]

$$f_{res,FC} = \frac{1}{2\pi}\sqrt{\frac{2.4^2 E_Y \varepsilon_r}{\rho R^2} - \frac{\epsilon_0}{0.813 \rho t g^3} V_G^2} \tag{7.3}$$

$$f_{res,DC} = \frac{1}{2\pi}\sqrt{\frac{\pi^2 E_Y \varepsilon}{\rho L^2} - \frac{16 \epsilon_0}{15 \rho t g^3} V_G^2} \tag{7.4}$$

where $E_Y$, $t$, and $\rho$ are, respectively, the Youngs' modulus, thickness, and mass density of the 2D material, $g$ is the gap size (distance from the 2D layer to the gate), and $\epsilon_0$ is the vacuum permittivity. For the fully clamped case, $R$ is the radius of the circular device, and $\varepsilon_r$ is the radial strain in the membrane. For the doubly clamped case, $L$ is the length of the device, and $\varepsilon$ is the axial strain in the membrane.

In both cases, contributions from electrostatic tensioning and capacitive softening are present. Specifically, electrostatic tensioning is represented by the first term under each square root, in which the effect of $V_G$ is implicit through modulating the strain in the device; capacitive softening is represented by the second term under each square root, in which the effect of $V_G$ is explicit. It is interesting to note that NEMS resonators with a higher initial tension (and thus frequency) generally exhibit smaller tunability (in frequency ratio) by $V_G$, because the $V_G$-induced strain is relatively small compared with the large initial strain. Therefore, by releasing the initial strain (for example, using thermal annealing), a much wider gate tuning range may be achieved.[57]

Device strain in 2D resonators can be induced via electrostatic tensioning (Figure 7.4a,b,c,d), thermal expansion (Figure 7.4b,e), external actuators (Figure 7.4f), pressure difference (Figure 7.4g), substrate bending,[178] and so on. In one example, the frequency of a $MoS_2$/graphene HS could be tuned up to 370% with gate voltage (Figure 7.4b).[179] In other examples,[180,181] the resonance frequencies of $MoS_2$ and graphene resonators have been tuned by temperature, and the direction of frequency tuning depends on the device structure and thermal expansion coefficient of the materials. A thermal hysteresis effect has also been exploited to realize frequency-reconfigurable resonators.[182] Additionally, using comb-drive actuators, the resonance frequencies of $MoS_2$, graphene, and h-BN resonators have been tuned (Figure 7.4f).[183]

# 8. QUALITY FACTOR AND DAMPING

Dissipation is important for MEMS/NEMS devices. In most cases, the rate of the dissipation can be characterized by the quality factor $Q$, which describes the ratio between the energy stored in the resonator and the energy dissipated during each vibration cycle:

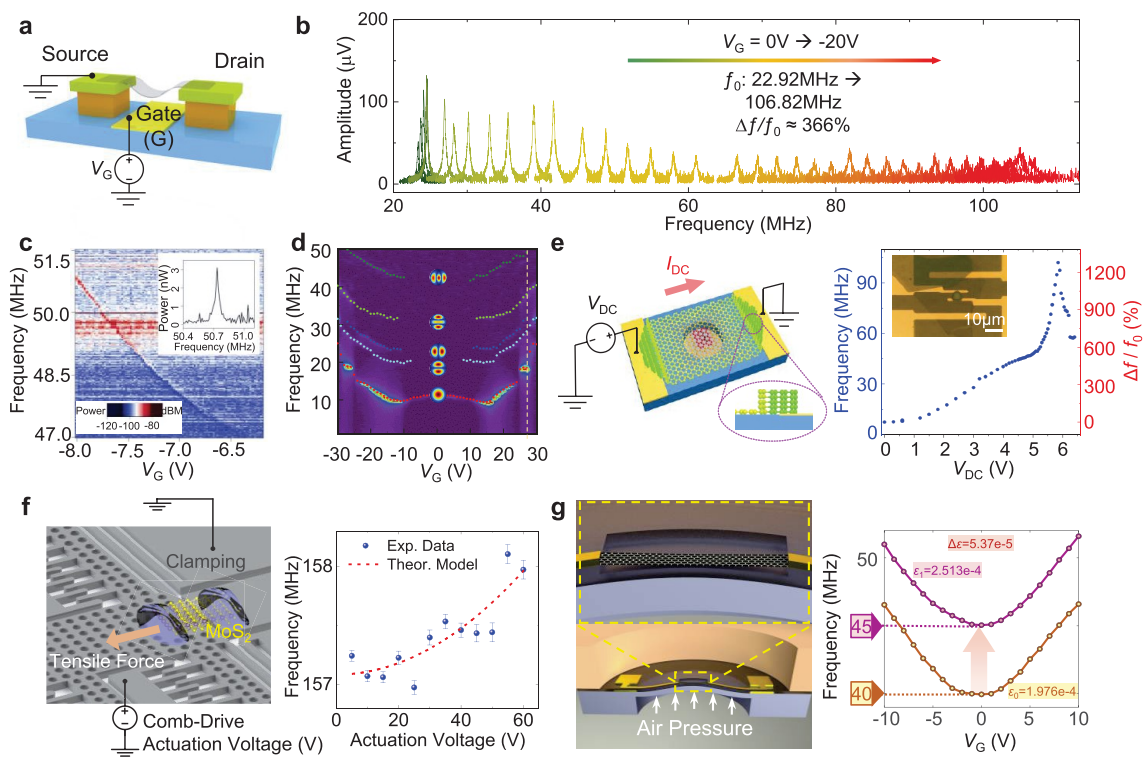

**Figure 7.4.** Frequency tuning in 2D NEMS resonators. (a) Schematic illustration of a doubly clamped 2D resonator, with deformation and strain induced by DC gate voltage. (b) Gate tuning of the resonance frequency for a 2D HS resonator. Reprinted in part with permission from ref 179. Copyright 2021 American Chemical Society. (c) Color plot of frequency tuning via DC gate voltage in a graphene NEMS oscillator. Reprinted in part with permission from ref 59. Copyright 2013 Nature Publishing Group. (d) Multimode resonances frequency tuning in a black P resonator. Reprinted in part with permission from ref 136. Copyright 2016 American Chemical Society. (e) Schematic of Joule heating in a fully clamped 2D resonator, and frequency tuning of a graphene resonator (inset) using Joule heating. Reprinted in part with permission from ref 258, copyright 2018 American Chemical Society, and ref 181, copyright 2018 Institute of Electrical and Electronics Engineers. (f) Illustration of a comb-drive actuator controlling strain in a doubly clamped $MoS_2$ resonator, and the measured frequency tuning data at different comb-drive actuation voltages. Reprinted in part with permission from ref 183. Copyright 2021 Wiley-Blackwell. (g) Schematic illustration for tuning the strain in a graphene resonator by deforming the membrane using pressure difference. Reprinted in part with permission from ref 245. Copyright 2021 Institute of Physics.

$$Q = \frac{\omega_0}{\gamma} = \frac{f_{res}}{BW} = 2\pi \frac{E_{st}}{E_{diss}} \quad (8.1)$$

where $f_{res}$ is the resonance frequency, $\gamma = \omega_0/Q$ (the 2nd term in eq 4.1, representing linear damping), BW = $\gamma/2\pi$ is the so-called "resonance bandwidth", which refers to the full width of the peak measured at half of the maximum power, and $E_{st}$ and $E_{diss}$ are respectively the energy stored in the resonator and the energy dissipated per cycle of vibration.

A larger quality factor corresponds to slower energy dissipation and a sharper resonance peak, and is generally desirable for device applications: It can lead to better frequency selectivity as well as a smaller power required to sustain the oscillation. With high-Q MEMS/NEMS resonators, researchers have demonstrated ultrasensitive transducers, narrow-passband and low-insertion-loss RF filters, self-sustained oscillators, etc.[184−187]

Both extrinsic and intrinsic dissipation mechanisms can exist in NEMS resonators.[188] Extrinsic dissipation sources can include air/fluid damping,[189] clamping loss, etc., while intrinsic dissipation mechanisms can include phonons in the resonator and the clamps,[190] surface-induced loss,[191−193] material-defect-induced dissipation,[194] thermoelastic damping,[195−197] etc. All of these need to be optimized to achieve high-Q resonators. A microscopic description of different dissipation mechanism is given in ref 190.

### 8.1. Quality Factors in 1D NEMS Resonators.
The quality factor in 1D NEMS resonators has been improving significantly over the years. Early measurements of CNT resonators at room temperature[11,198] typically showed Q values below 2000. A number of approaches have since been adopted to improve Q in 1D resonators. Cooling: By cooling doubly clamped CNT resonators down to millikelvin temperature under high vacuum, the effects from phonon and air damping can be minimized, and Q values exceeding 5 million have been measured using ultralow-noise measurement instruments (Figure 8.1a,b).[48,104] In one work,[104] several effects appearing in CNTs have been carefully studied and controlled, which are important for the observed high Q: (i) electromechanical damping suppressed in the Fabry−Pérot regime and (ii) apparent enhancement of Q through the reduction of the phase noise. Strain: Strain tuning of Q has been demonstrated in NW resonators by bending the substrate, and Q up to 390,000 at room temperature have been demonstrated (Figure 8.1c).[199] It has also been reported that an external DC voltage can increase the axial strain in cantilever CNT resonators, which results in "soft clamping", and thus enhances the Q values.[200] Limiting amplitude: Large vibration amplitudes can lead to nonlinear effects such as nonlinear damping,[101] and thus reduce Q values. Therefore, limiting the vibration amplitude, such as by reducing the AC electrical driving, can sometimes also improve $Q$ values in CNT resonators.[201]

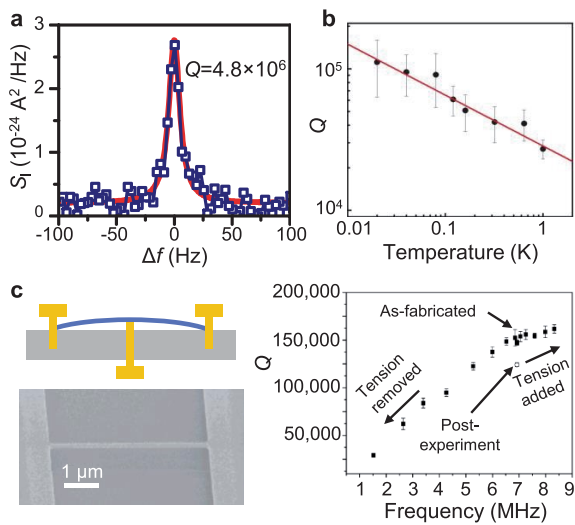

**Figure 8.1.** Controlling and enhancing $Q$ in 1D NEMS resonators. (a) Resonance spectrum of a CNT resonator showing $Q$ close to 5 million. Reprinted in part with permission from ref 104. Copyright 2014 Nature Publishing Group. (b) Temperature dependence of $Q$ in a CNT resonator. Reprinted in part with permission from ref 40. Copyright 2009 American Chemical Society. (c) Strain tuning of $Q$ and $f$ in a silicon nitride NW resonator by substrate bending. Reprinted in part with permission from ref 199. Copyright 2007 American Chemical Society.

**8.2. Quality Factors in 2D NEMS Resonators.** Graphene resonators are arguably the most studied type of 2D NEMS resonators, and their quality factors can be affected by a number of factors and mechanisms. <u>Size</u>: A study of graphene mechanical resonators with different diameters shows that $Q$ scales with device size (smaller devices have lower $Q$s).[123] <u>Stacking</u>: Significant difference in $Q$ is found among monolayer, twisted bilayer, and Bernal-stacked bilayer graphene resonators,[202] suggesting the vdW interaction between layers, or more specifically the interlayer friction, can affect the energy dissipation in multilayer 2D NEMS resonators. <u>Clamping geometry</u>: Unlike 1D CNT resonators, doubly clamped 2D resonators have free edges which can also lead to dissipation, and both theoretical[203] and experimental[60,204] studies show that resonators with free edges generally exhibit lower $Q$ compared with those without free edges. <u>Amplitude</u>: Amplitude of motion can also have strong effects on the nonlinear damping, and a graphene with quality factor reaching 100,000 has been demonstrated by exploiting the nonlinear nature of damping through controlling the driving amplitude.[101] <u>Temperature</u>: By cooling the device to 15 mK and by reducing the vibration amplitude, a graphene NEMS resonator with $Q$ over 1 million has been experimentally demonstrated (Figure 8.2a).[142]

Besides graphene, 2D NEMS resonators based on other 2D materials have also been studied to understand and control dissipation in these atomically thin mechanical structures (Figure 8.2b–g). A theoretical study suggests that the $Q$ of $MoS_2$ resonators could be intrinsically higher than that of graphene resonators due to a larger phonon gap.[129] Pressure-dependent studies of $MoS_2$ resonators suggest that free-molecule-flow damping plays an important role in 2D resonators,[189] leading to a different pressure dependence of $Q$ in 2D resonators than in most 3D MEMS resonators. The temperature dependence of the quality factor of $MoS_2$ resonators has been compared to models from different dissipation mechanisms (Figure 8.2c).[205] The $Q$ dependence on device tension (Figure 8.2e)[183,206] and DC gate voltage has also been experimentally studied (Figure 8.2f).[63,207] Furthermore, the trend of tuning $Q$ using a DC gate voltage is dependent on the device geometry (Figure 8.2g).[208] Increasing the AC drive voltage generally decreases the $Q$ due to the increased vibration amplitude (Figure 8.2d).[208,101]

Heterostructures can be used to enhance quality factors in 2D resonators. For example, resistive loss through Joule heating can be important for NEMS resonators using electrical readout, especially at low temperatures. Through adding a low resistivity $NbSe_2$ layer between two graphene layers in a graphene/$NbSe_2$/graphene resonator,[154] such electrical loss is reduced and the mechanical $Q$ in the device can reach up to 245,000.

## 9. NONLINEARITY AND DYNAMIC RANGE

**9.1. Nonlinearity in NEMS Resonators.** As shown in eq 4.1, when the vibration amplitude increases, the nonlinear terms of the restoring and damping forces become non-negligible and manifest in the device dynamics. Depending on which nonlinear term is the dominant correction to the linear equation of motion, the device can be described using the Duffing equation (including a cubic term in the spring constant):[63]

$$\ddot{z} + \gamma\dot{z} + \omega_0^2 z + \alpha z^3 = \frac{F(t)}{m} \tag{9.1}$$

or the van der Pol equation (including a cubic term in the damping coefficient):[101]

$$\ddot{z} + \gamma\dot{z} + \omega_0^2 z + \eta z^2 \dot{z} = \frac{F(t)}{m} \tag{9.2}$$

leading to different nonlinear responses (Figure 9.1a,b,c). Nonlinearities can be observed even at relatively small vibration amplitudes. Therefore, nonlinear responses have important implications for device dynamics. They can also be used to extract device parameters such as Young's modulus.[209]

There can be many different sources for spring nonlinearities, and they can generally be categorized based on whether they increase or decrease the resonance frequency upon increasing the driven displacement (hardening or softening; Figure 9.1f shows examples of both). It is often true that several types of nonlinearities exist in a NEMS resonator. Since the electrostatic gate force tunes the static tension in the membrane, and such force tends to break the symmetry of the mode shape, it can be employed to increase the contribution to the nonlinear Duffing force associated with softening of the resonant mode. This can sometimes lead to the ability to realize zero effective nonlinearity for not too large displacement or to control the resonator in either a hardening or softening regime.[125,210,211] In addition, while the conventional wisdom for resonant sensors is to avoid nonlinear responses to preserve the linearity of the output, researchers have found that the response of NEMS resonant sensors can be enriched by exploiting the nonlinear effects.[212,213] All this is possible because the suspended nanomaterials can withstand a very high level of strain, allowing them to vibrate with very large amplitudes (compared with device size) and thus operate in deep nonlinear regimes; this feature is not readily available in bulk materials (such as Si, which is commonly used for MEMS resonators) that tend to fail at moderate strains.

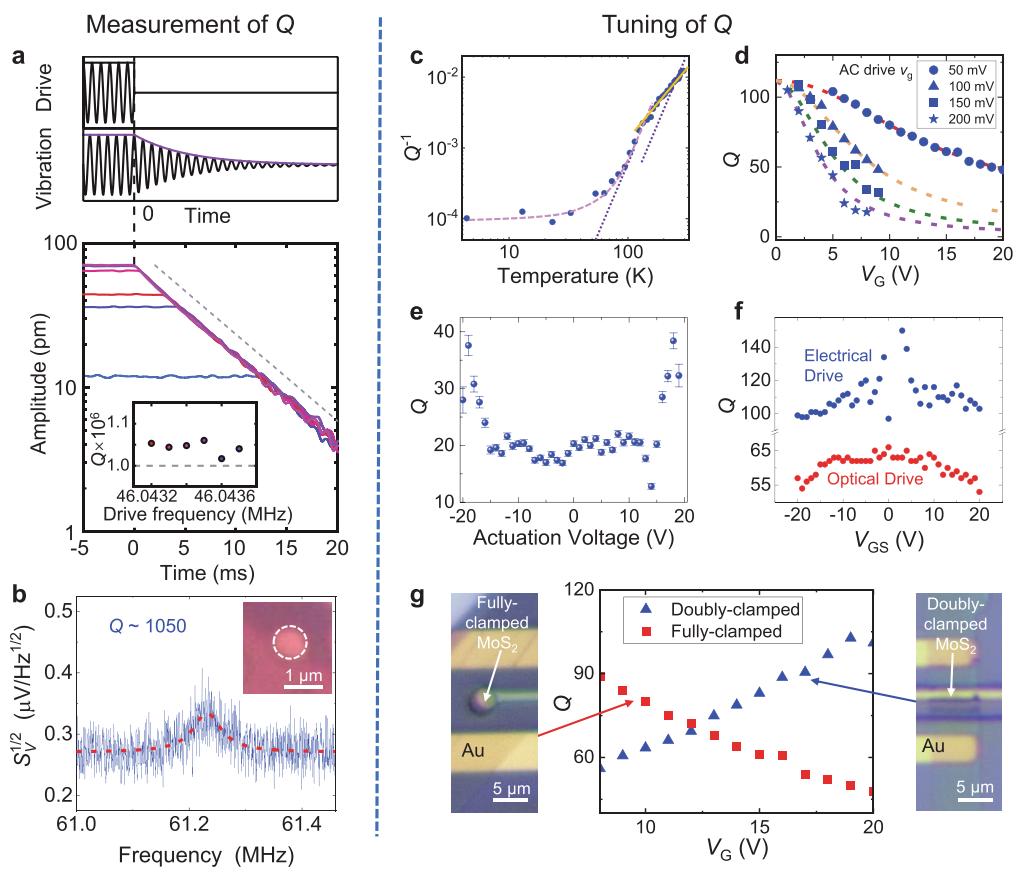

**Figure 8.2. Controlling and enhancing Q in 2D NEMS resonators. (a)** A graphene resonators measured at 15 mK, showing Q exceeding 1 million. Reprinted in part with permission from ref 142. Copyright 2021 Springer Nature. **(b)** Thermomechanical resonance of a 2D MoS₂ resonator measured at room temperature, showing Q exceeding 1000. Reprinted in part with permission from ref 63. Copyright 2018 The Authors, some rights reserved; exclusive licensee AAAS. Distributed under a CC BY-NC 4.0 license http://creativecommons.org/licenses/by-nc/4.0/. **(c)** Temperature dependence of Q for a 2D MoS₂ resonator. Reprinted in part with permission under a Creative Commons (CC-BY-NC-ND) License from ref 205. Copyright 2017 Nature Publishing Group. **(d)** Dependence of Q on DC and AC gate voltages, for 2D MoS₂ NEMS resonators. Reprinted in part with permission from ref 208. Copyright 2022 American Chemical Society. **(e)** Q vs comb-drive actuation voltage for 2D MoS₂ resonator mounted on a comb-drive. Reprinted in part with permission from ref 183. Copyright 2021 Wiley-Blackwell. **(f)** Gate tuning of Q for MoS₂ resonators. Reprinted in part with permission from ref 63. Copyright 2018 The Authors, some rights reserved; exclusive licensee AAAS. Distributed under a CC BY-NC 4.0 license http://creativecommons.org/licenses/by-nc/4.0/. **(g)** Different effect of tuning Q for fully clamped and doubly clamped 2D MoS₂ NEMS resonators using DC gate voltage. Reprinted in part with permission from ref 208. Copyright 2022 American Chemical Society.

Here we take a closer look at CNT resonators: The extreme aspect ratio of CNT and thus the large force response has important consequences for the dynamics of nanotube devices.[95,214] For instance, a typical nanotube with a length of 1 $\mu$m attains a room-temperature thermal motion of $z_{th}$ = 2.5 nm, a large displacement that likely drives the device into the nonlinear regime. The fluctuating amplitude of the thermal motion translates into fluctuations of the resonance frequency due to the Duffing nonlinearity, causing an apparent broadening of the resonance line width and an apparent reduction of the Q factor (when it is extracted from spectral measurements).[97] The spectrum can be further broadened by the thermal motion of the other modes through the dispersive mode coupling (Section 10.1).[174,215] Such spectral broadening is not related to energy loss but to decoherence and frequency noise.[104,216] Recent measurements also indicate that the thermal energy could concentrate in one mode, disperse into a second mode, and then return, reminiscent of the Fermi–Pasta–Ulam–Tsingou behavior.[217] Therefore, the thermal noise combined with the Duffing nonlinearity and the mode coupling leads to rich dynamics in CNT resonators and deserves further studies in the future.

## 9.2. Linear Dynamic Range.

Another important device characteristic that depends on both the thermal noise and the nonlinearity is the linear DR. It describes the range of amplitude in which the vibration response is linear, i.e., the motional amplitude of the device (output) is proportional to the driving amplitude (input). Such proportionality is important for sensing and signal transduction, and a larger linear DR is generally desirable for such devices. The linear DR is defined as the ratio of the amplitude at the onset of nonlinearity (beyond which the vibration is considered nonlinear) to the standard deviation of the thermomechanical noise at resonance integrated over the measurement bandwidth, i.e., the range in which linear resonance response can be obtained (Figure 9.1a). In a resonator with Duffing nonlinearity (which is by far the most common type of nonlinearity observed), its linear DR can be expressed as[95]

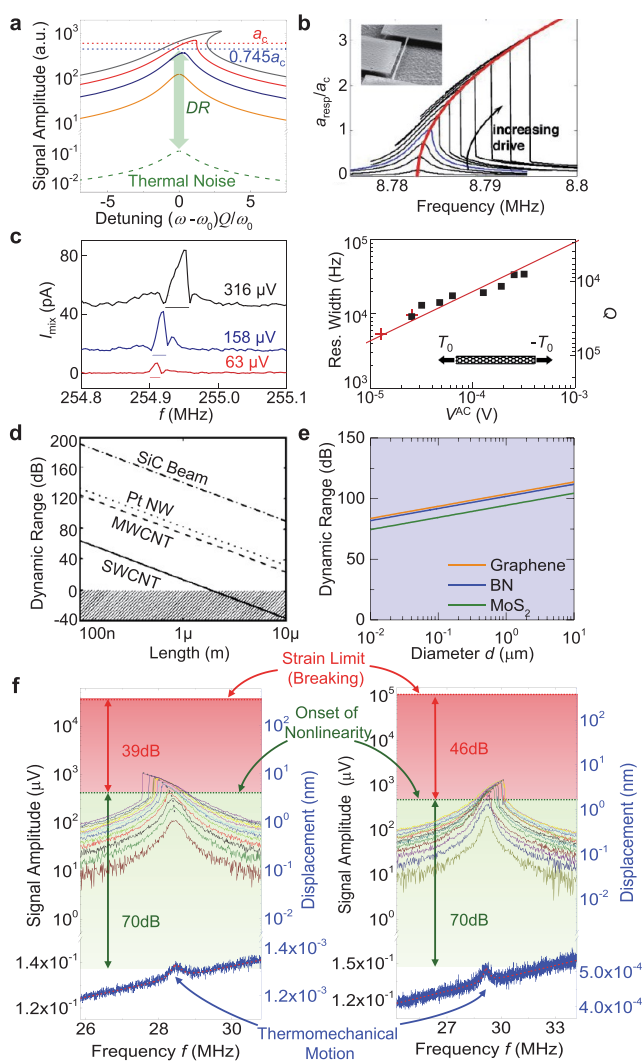

**Figure 9.1.** Nonlinearity and DR in 1D and 2D NEMS resonators. (a) Definition of DR in a NEMS resonator, from thermomechanical resonance to onset of nonlinearity as vibration amplitude increases. Reprinted in part with permission from ref 218. Copyright 2014 American Institute of Physics. (b) Resonance spectra of the doubly clamped SiC NW resonator with increasing drive, showing Duffing nonlinearity. Reprinted in part with permission from ref 172. Copyright 2006 American Institute of Physics. (c) Nonlinear damping in a CNT resonator. Reprinted in part with permission from ref 101. Copyright 2011 Nature Publishing Group. (d) Calculated DR for doubly clamped 1D NEMS resonators with different lengths, including CNT and NW resonators. Reprinted in part with permission from ref 95. Copyright 2005 American Institute of Physics. (e) Calculated DR for fully clamped circular 2D NEMS resonators at 300 K. Reprinted in part with permission from ref 218. Copyright 2014 American Institute of Physics. (f) Softening and hardening nonlinearities measured in monolayer and 3-layer 2D MoS₂ NEMS resonators as well as DRs. Reprinted in part with permission from ref 63. Copyright 2018 The Authors, some rights reserved; exclusive licensee AAAS. Distributed under a CC BY-NC 4.0 license http://creativecommons.org/licenses/by-nc/4.0/.

$$DR = 20\log\left(\frac{0.745a_c}{\sqrt{2S_{x,th}\Delta f}}\right) \qquad (9.3)$$

where $a_c$ is the critical amplitude (the onset of bistability, and $0.745a_c$ is 1 dB below the critical amplitude, often called the 1 dB

compression point), $S_{x,th}^{1/2}$ is the on-resonance thermomechanical noise spectral density, and $\Delta f$ is the measurement bandwidth. As devices scale down, the DR exhibits different trends for 1D and 2D NEMS resonators.

*9.2.1. DR in 1D NEMS Resonators.* The device aspect ratio can have a large effect on the linear DR. Calculations[95] show that as the length of the NW or CNT increases, the DR decreases (Figure 9.1d). When the device is sufficiently long or thin, the resonator could be even intrinsically nonlinear (Figure 9.1b), showing no linear region of operation. Experimentally, VLS synthesized Si NW cantilevers with a wide range of dimensions have been studied, showing DR ranging from 53 to 90 dB.[83] Studies of doubly clamped SiC NW resonator show that the DR can be tuned from 69 to 73 dB by applying a gate voltage,[172] in good agreement with the theoretical model.

*9.2.2. DR in 2D NEMS Resonators.* 2D NEMS resonators are found to exhibit broader DR compared with 1D NEMS resonators. Interestingly, a theoretical analysis of 2D drumhead resonators[218] shows that DR increases with device diameter (Figure 9.1e), in striking contrast to 1D resonators (Figure 9.1d). It also shows that tensile strain in 2D NEMS resonators can lead to a larger DR. Such broad DR has been observed in mono-, bi-, and trilayer MoS₂ drumhead resonators, with measured linear DR up to 70 dB (Figure 9.1f),[63] showing excellent agreement with the theory. DR tuning has also been experimentally demonstrated for 2D graphene resonators, in which the DR can be enhanced by up to 25 dB using gate voltage-induced strain.[165]

## 10. MECHANICAL MODE COUPLING IN LOW-DIMENSIONAL RESONATORS

Mechanical coupling refers to the interaction between two (or more) mechanical eigenmodes—they can be the modes in either different structures[219−226] or within the same vibrating structure.[171,227−231] An easy way to visualize such effects is to think about two spring-mass resonators connected by a coupling spring (Figure 10.1), whose equations of motion can be written as

$$m_1\ddot{z}_1 + \gamma_1\dot{z}_1 + k_1 z_1 - k_c(z_2 - z_1) = F_1(t)$$

$$m_2\ddot{z}_2 + \gamma_2\dot{z}_2 + k_2 z_2 + k_c(z_2 - z_1) = F_2(t) \qquad (10.1)$$

where $z_i$ is the displacement of the $i^{th}$ resonator ($i = 1$ or $2$). It can be clearly seen that with the presence of the coupling spring $k_c$, the motion of the first resonator depends on the second one, and vice versa. In two resonators situated in different places, the coupling $k_c$ can occur via the transfer of elastic energy between different resonators.[219−221,226] Meanwhile, within a multimode resonator, the coupling between different modes can arise through the stress in the resonator or a force gradient.

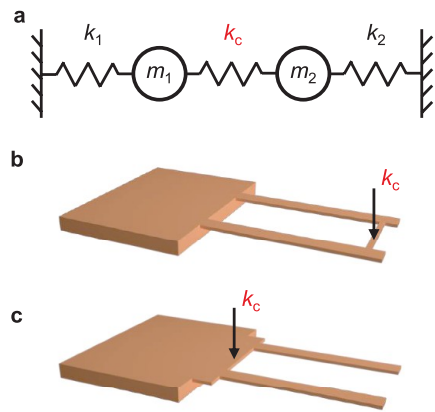

**Figure 10.1.** Coupled resonators. (a) Schematic illustration of two coupled modes in a spring-mass model. (b) Illustration of two singly clamped beam resonators coupled through a coupling beam. (c) Illustration of two singly clamped beam resonators coupled through a common clamping edge. In both (b) and (c), the origin of the coupling spring term $k_c$ is indicated.

Mechanical coupling in micro- and nanoscale resonant devices can give rise to a plethora of interesting physical phenomena such as mechanically induced transparency[232−234] and can be leveraged to realize unique device functions like transferring and storage of information in both the classical and quantum domains.[235−237] Due to their ultrasmall resonant mass and superb frequency tunability as well as the unique physical properties offered by various 1D and 2D nanomaterials, mechanical coupling in low-dimensional NEMS resonators is of particular interest to researchers for exploring exquisite physical processes and demonstrating various device applications.

It is important to note that eq 10.1 describes the simplest case, i.e., linear resonant coupling. Its result is simply converting the original set of eigenmodes into another set, i.e., the "coupled modes". When additional terms (such as higher-order terms and periodic modulations to stiffness or damping) are included, one can have more interesting resonant coupling, which can be further categorized into nonlinear coupling and parametric coupling. For more detailed and in-depth discussion, we refer the readers to a more dedicated review.[190] Here we use a phenomenological and simplified approach to introduce mode coupling in 1D and 2D NEMS resonators.

**10.1. Coupling through Nonlinearity.** The high mechanical strength of 2D materials allows 2D resonators to be driven at very high amplitudes, often more than the membrane thickness, leading to strong nonlinear effects.[212,213] In addition, the static tension tunes the frequency of different resonant modes. Tension-tunability of 2D membranes allows manipulating individual mode frequencies, nonlinearities, and mode coupling, which is not readily available in most MEMS devices.

Coupled modes affect each other through tension. The amplitude of a particular mode can modify the resonant frequencies of all the modes that are coupled through tension.[238] This dispersive coupling does not require any condition on the resonance frequency ratio of the two modes.

Another scenario arises when the modes have commensurate resonance frequencies. Due to modal nonlinearities, the response of a resonator is modified and includes higher tones that are integer multiples of the excitation frequency; the high-frequency mode is driven by a harmonic of the low-frequency mode, and the low-frequency mode is driven by a subharmonic of the high-frequency mode (Figure 10.2a).[94,132] This phenomenon is known as "internal resonance", which enables the efficient exchange of energy between the modes at different eigenfrequencies. It can lead to peculiar resonant features such as the frequency pinning of the driven mode (Figure 10.2b)[132,239] and allows the study of nonlinear damping.[240]

Similar phenomena have also been observed in CNT resonators. By applying a static force within the nanotube and sweeping the resonant frequencies, one can tune the different modes in and out of nonlinear resonant coupling, i.e., internal resonance, as the ratio of the eigenfrequencies of the two modes is close to an integer $n$ or a rational number $n/m$, e.g., $1/3$, $2$, or $2/3$ (Figure 10.3).[94]

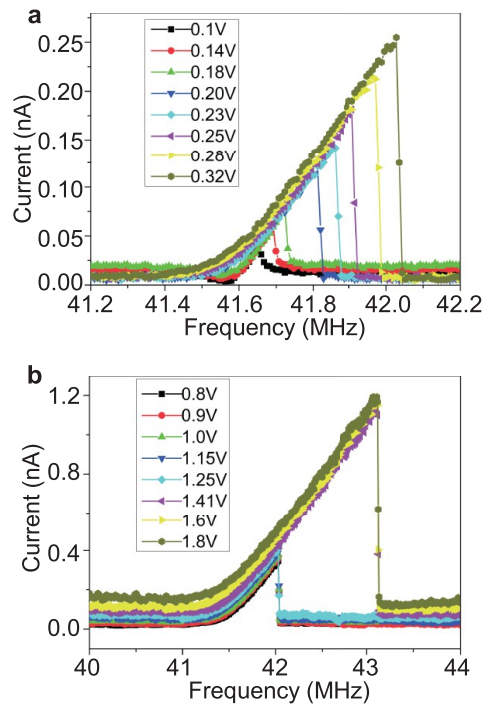

**Figure 10.2.** Internal resonance and frequency pinning in an $MoS_2$ NEMS resonator. (a) The frequency response of the device with positive Duffing nonlinearity. (b) Pinning of the jump-down frequency when the internal resonance occurs in the multistable regime. Reprinted in part with permission from ref 132. Copyright 2015 American Institute of Physics.

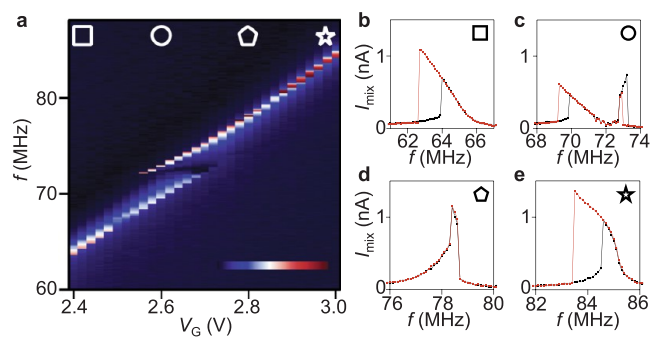

**Figure 10.3.** Internal resonance in a CNT resonator. (a) Nonlinear resonant coupling revealed by measuring the oscillation amplitude of a single mode with the two-source current mixing method[11] as a function of gate voltage $V_G$. The measured current ranges from 0 (black) to 1 nA (red). An avoided crossing is observed when the frequency is half that of a higher-order mode. Reprinted in part with permission from ref 94. Copyright 2012 American Physical society. (b–e) Response of the vibration amplitude (which is roughly proportional to the current) to an oscillating force at different gate voltages indicated by the symbols in (a). Black and red lines are sweeps with increasing and decreasing frequency, respectively. The line shapes are highly irregular due to an interplay of nonlinear resonant coupling and nonlinear Duffing response oscillations. Reprinted in part with permission from ref 94. Copyright 2012 American Physical society.

**10.2. Coupling through Parametric Pumping.** The coupling between two modes can be turned on by applying a periodic signal at neither their frequencies. Such signal is often called the "pump", with frequency $\omega_p$, which parametrically drives the mode coupling. The coupled equations of motion then take the form:[231,241,242]

$$\ddot{z}_1 + \gamma_1 \dot{z}_1 + [\omega_1^2 + \Gamma_1 \cos(\omega_p t)]z_1 + \Lambda \cos(\omega_p t)z_2 = \frac{F_1}{m_1}\cos(\omega_d t + \phi_1)$$

$$\ddot{z}_2 + \gamma_2 \dot{z}_2 + [\omega_2^2 + \Gamma_2 \cos(\omega_p t)]z_2 + \Lambda \cos(\omega_p t)z_1 = \frac{F_2}{m_2}\cos(\omega_d t + \phi_2)$$

(10.2)

The terms containing $\Gamma$ and $\Lambda$ are controlled by the parametric pump signal and are responsible for sinusoidal modulation of the stiffness of the mode by stretching of the nanomaterial, i.e., the resonator body. This periodic stretching can be realized, for example, using electrical gating[227,231] or photothermal heating.[171,228] The terms proportional to $\Gamma_i$ are responsible for the parametric amplification effect mentioned in Section 4.2.2,[98,114,137,243,244] but for the case discussed here they are usually far off-resonance, and produce no effect. The pump signal $\Lambda \cos(\omega_p t)$ controls the coupling between the two modes when $\omega_p = |\omega_1 - \omega_2|$. In such a process with $\omega_1 < \omega_2$, a quantum with energy $\hbar\omega_p$ and a quantum with energy $\hbar\omega_1$ are annihilated while a quantum with energy $\hbar\omega_2$ is created, and vice versa (Figure 10.4a).

Besides the requirement of $\omega_p = |\omega_1 - \omega_2|$, the coupling strength is determined by the magnitude of $\Lambda$. When the rate of transfer of phonons between the modes ($\propto \Lambda$) is larger than the rate of decay of phonons in each mode ($\propto \gamma_1 + \gamma_2$), the system is said to be in the strong coupling regime, where each of the two resonances splits.[241] Such splitting is observed in many 1D and 2D NEMS resonators (Figure 10.4b,c).[171,219,227–231] The amount of splitting (denoted as $2g$) increases with $\Lambda$ and can be controlled by modulating the strength of the pump signal (Figure 10.4d,e),[227,229,231] tuning the tension using a DC gate bias,[227,231] heating the resonator using a laser beam,[171] or even bending the substrate.[245]

Transfer of energy between the modes can also occur[241] when $\omega_p = |\omega_1 - \omega_2|/n$, for integer values of $n \geq 2$.[171,227–231] An example of this phenomenon in 2D NEMS resonators is shown in Figure 10.4f. When pumped at the sum $\omega_p = \omega_1 + \omega_2$, the modes experience mutual driving. This case is not energy conserving and can result in strong amplification of both amplitudes.

# 11. NEMS RESONATORS FOR SENSING

NEMS resonators based on low-dimensional nanomaterials possess many advantages, such as ultrasmall size (down to just one or few atoms in at least one dimension), ultrahigh aspect ratio (exceeding $10^6$ in some CNT resonators), ultralow mass (down to attogram level; below 100,000 carbon atoms for the shortest CNT), high resonance frequencies (as high as >10 GHz), ultralow power consumption (down to subpicowatt), and large frequency tuning ranges (exceeding 2000%). Thanks to these advantages, NEMS devices have demonstrated exciting potentials toward a number of applications, with sensing applications explored most extensively.

**11.1. Mass Sensing.** One of the most notable applications for NEMS resonators is mass sensing. As can be seen from the expression of resonance frequency for a simple spring-mass resonator (eq 7.1,) any change in the resonator mass impacts the resonance frequency. The smaller the initial resonator mass, the larger the frequency shift is due to a certain added mass. Thanks to their ultrasmall mass, NEMS resonators make exceptional mass sensors.

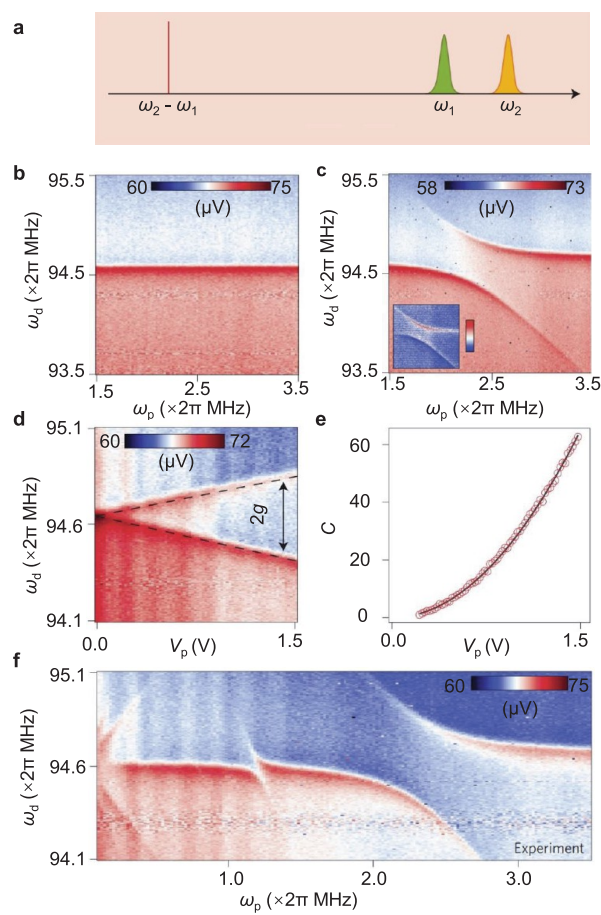

**Figure 10.4.** Parametric coupling in NEMS resonators. (a) Illustration of a red-detuned pump alongside the two coupled modes in the frequency domain. (b, c) The response of mode 1 as a function of the red-pump detuning when $V_p = 0\,V$ (b) and $V_p = 1.5\,V$ (c) in a graphene drumhead resonator. For nonzero pump amplitudes, mode 1 is seen to split in the vicinity of $\omega_p \approx |\omega_1 - \omega_2|$. Inset, the response detected at $\omega_d + \omega_p$. Energy transfer to the second mode is seen when $\omega_d + \omega_p \approx \omega_2$. (d) Response of mode 1 at fine intervals of the pump voltage. (e) Cooperativity, a figure of merit that quantifies the magnitude of intermodal coupling vs pump amplitudes. The solid line is a quadratic fit of the data (circles) to the equation $C = \alpha V_p^2$. (f) Higher-order parametric coupling in NEMS resonators: Mode 1 response as a function of the pump detuning over a larger frequency range in a graphene drumhead resonator. Apart from normal mode splitting at $\omega_p \approx \Delta\omega = |\omega_1 - \omega_2|$, an extra splitting at $\omega_p \approx (\Delta\omega)/2$ can be seen, suggesting the onset of higher-order intermodal coupling. (a–f) Reprinted in part with permission from ref 227. Copyright 2016 Nature Publishing Group.

Quantitatively, when an added mass of amount $\Delta m$ lands on a resonator of mass $M$, the resonance frequency $f_n$ (for the nth resonance mode) will be shifted by $\Delta f_n$ according to[246]

$$\frac{\Delta f_n}{f_n} = \frac{-\Delta m}{M} \frac{\phi_n(x)^2}{\alpha_n}$$

(11.1)

where $\phi_n(x)$ denotes the relative amplitude (ratio to the maximum displacement point on the resonator) for the nth mode at position $x$, the position of adsorption for the added mass $\Delta m$. The numerical factor $\alpha_n$, which is on the order of unity, depends on the specific mode. Again, it can be seen that the fractional frequency shift, $\Delta f_n/f_n$ is proportional to the fractional mass change $\Delta m/M$. Therefore, a smaller resonator mass $M$ means that for the same amount of resonance frequency shift, a smaller added mass $\Delta m$ can be detected.

To achieve the ultimate mass sensitivity, researchers have been striving to build ever smaller NEMS resonators. For example, SiC NW resonators have been demonstrated to function as highly sensitive mass sensors (Figure 11.1a)[247] with a mass resolution down to 7 zg (1 zg = $10^{-21}$ g), corresponding to ∼30 xenon atoms.[248] Later, using CNT resonators which have smaller masses than NW ones, the detection of small amount of atoms such as platinum (Pt, 0.324 zg),[56] xenon (Xe, 0.218 zg),[71] chromium (Cr, 0.1 zg),[249] and argon (Ar, 0.066 zg)[71] have been demonstrated (Figure 11.1b). Further, by using an ∼100 nm short CNT resonator (even smaller mass) and annealing it to reduce frequency fluctuations, a mass resolution on the order of a single proton (1.7 yg, 1 yg = 0.001 zg) has been achieved (Figure 11.1c).[250]

Besides 1D NEMS resonators, mass sensing has also been demonstrated in 2D resonators. For example, graphene resonators have been used for sensing pentacene molecules (Figure 11.1f), and a mass sensitivity of ∼2 zg has been achieved.[57] While their mass sensitivity is not as competitive as that of CNT resonators (given the larger resonator masses), the large surface area makes 2D NEMS resonators particularly promising for capturing low-density, low-concentration incoming particle fluxes, which can be a practical advantage in many applications.[251]

The superb mass resolution of low-dimensional resonators allows researchers to study a number of exotic physical processes related to surface adsorption, such as the kinetics and fluctuations of adsorbed atoms over the device surface (Figure 11.1d),[46,247,252] the formation of monolayers of adsorbed atoms in low-dimensional solid, liquid, and vapor phases, and the associated phase transitions (Figure 11.1e)[41,42,46] as well as the adsorption of helium superfluid monolayers and the so-called layering transition.[47]

## 11.2. Force Sensing.
Low-dimensional NEMS make excellent force sensors. The ultimate force sensitivity is limited by the thermal force noise, which is given by the fluctuation−dissipation theorem:[91]

$$S_F = 4k_B T M_{eff}\gamma = 4k_B T\sqrt{M_{eff}k_{eff}}/Q \qquad (11.2)$$

where $k_B T$ is the thermal energy, $\gamma = \omega_0/Q$ is the damping coefficient, $M_{eff}$ is the eigenmode mass, $k_{eff}$ is the effective spring constant, and $Q$ is the quality factor. Therefore, the high flexibility (thus low spring constant), small masses, and high quality factors make low-dimensional resonators particularly advantageous for force sensing.

Similar to the quest for the ultimate mass sensitivity, researchers have been pushing the limit of force sensing with continued scaling of NEMS resonators. For example, NW resonators (including many with singly clamped geometry) have been extensively explored for sensing small forces.[253] Smaller in size and mass, CNT resonators can achieve force sensitivities as low as 4.3 zN/√Hz (1 zN = $10^{-21}$ N) when operating the device at low temperature (0.12 K).[254] Central to this result is lowering the contribution from the noise of the detection. This sensitivity can enable the detection of tiny electrical forces (Figure 11.2a).[91] Potentially, it could also allow coupling the nanotube motion to the spin degree of freedom of individual electrons[255] and protons (hydrogen atoms),[256] toward single-atom nuclear magnetic resonance and quantum spin-phonon physics.

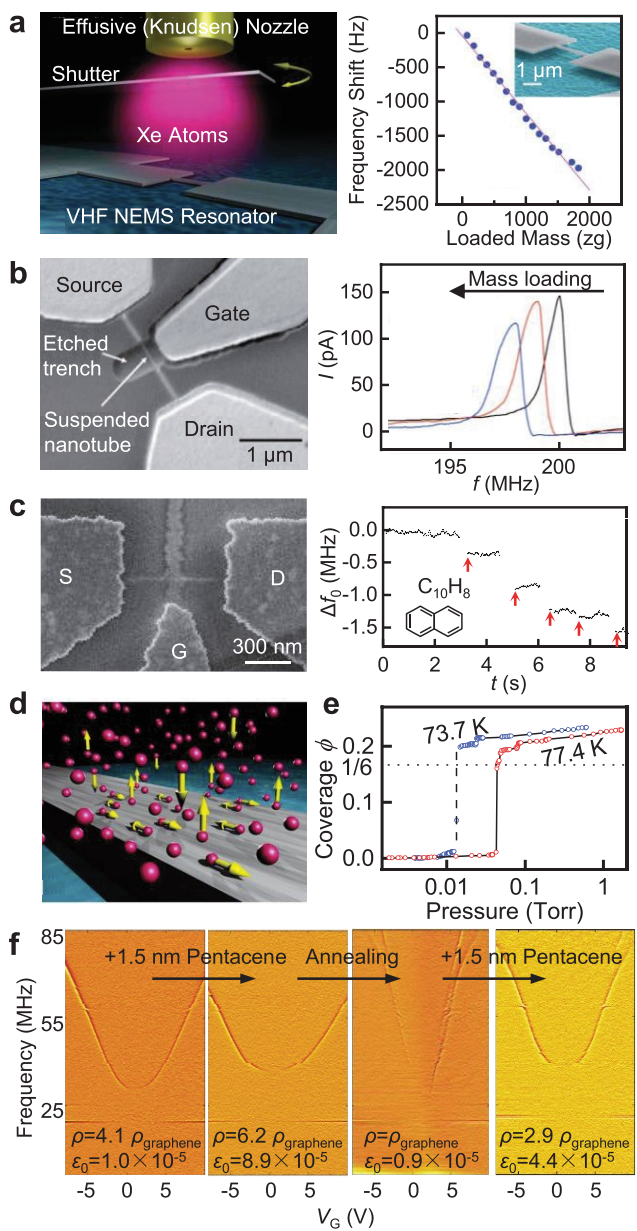

Figure 11.1. Mass sensing applications using NEMS resonators. (a) Schematic illustration of Xe atom mass sensing using a SiC NW resonator and the measured frequency response to mass loading, showing zg-level mass sensing. Reprinted in part with permission from ref 247. Copyright 2011 American Chemical Society. (b) SEM image and resonance curve during adsorption for a CNT resonator. Reprinted in part with permission from ref 71. Copyright 2008 American Chemical Society. (c) SEM image and mass sensing data for a CNT resonator, showing down to 1.7 yoctogram resolution. Reprinted in part with permission from ref 250. Copyright 2012 Nature Publishing Group. (d) Schematic illustration of fluctuation of Xe atoms (including adsorption, desorption, and surface diffusion) on a SiC NW resonator. Reprinted in part with permission from ref 247. Copyright 2011 American Chemical Society. (e) Low-dimensional phase transition of Kr atoms adsorbed on the surface of a CNT resonator, detected by monitoring the resonance frequency. Reprinted in part with permission from ref 42. Copyright 2010 American Association for the Advancement of Science. (f) Mass sensing behavior of graphene resonator. Reprinted in part with permission from ref 57. Copyright 2009 Nature Publishing Group.

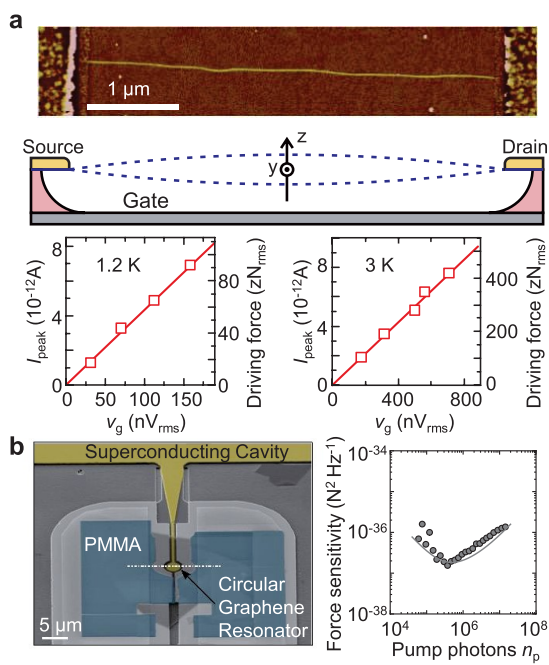

Figure 11.2. Force sensing applications of low-dimensional NEMS resonators. (a) AFM image of a 4 $\mu$m long CNT, schematic of the device, and the corresponding force sensing experimental results showing zN force sensing at 1.2 K and 3 K. Reprinted in part with permission from ref 91. Copyright 2013 Nature Publishing Group. (b) False-colored image of a multilayer graphene resonator and the force sensitivity under different numbers of pump photons. Reprinted in part with permission under a Creative Commons (CC BY) License from ref 77. Copyright 2016 Nature Publishing Group.

2D resonators have also been explored for sensitive force detection. For example, the first graphene resonator shows a force sensitivity down to 0.9 fN/$\sqrt{Hz}$ (1 fN = $10^{-15}$ Newton).[16] By coupling a graphene resonator to a superconducting cavity, a force sensitivity of 390 zN/$\sqrt{Hz}$ has been achieved (Figure 11.2b), together with a displacement sensitivity of 1.3 fm/$\sqrt{Hz}$.[77]

### 11.3. Other Sensing Applications.
Besides mass and force sensing, low-dimensional NEMS resonators have been explored for other sensing applications.

*11.3.1. Temperature Sensing.* Varying temperature will change not only the Young's modulus of the material[257] but also the tension in the suspended structure due to the differential thermal expansion of materials, both of which can lead to frequency shifts in response to changing temperatures. Different NEMS resonators exhibit different temperature-dependent behaviors. For example, $MoS_2$ resonators and h-BN resonators show negative temperature coefficients of frequency (TC$f$), defined as the relative frequency shift per temperature change. In recent experiments, $MoS_2$ resonators exhibit a TC$f$ of −0.396%/K from 293 to 315.5 K, and h-BN resonators show a TC$f$ of −0.285%/K from 266 to 414 K.[20,180] In contrast, graphene NEMS resonators exhibit a positive TC$f$ of 1.2%/K from room temperature to 1250 K,[258] and can operate up to 2650 K.[181] These devices could potentially be used as resonant temperature sensors.

*11.3.2. Light and Radiation Sensing.* With a rich collection of materials with different bandgaps, NEMS based on low-dimensional materials can be used for sensing photons in a wide range of wavelengths. Black P has a bandgap that varies from 0.3 to 2 eV depending on the thickness, thus black P resonators are capable of covering a wide range of wavelengths from visible light to infrared radiation. In one example, the device frequency shifted upon exposure to 785 nm light due to the photothermal effect, with a responsivity of $\mathcal{R}$ = −0.31 kHz/μW.[259] $MoS_2$ resonators have been demonstrated to show different responsivities for lights of different wavelengths (405, 532, and 633 nm).[131,180] In addition, photons with even higher energy, such as $\gamma$-ray radiation, can also be detected using NEMS resonators. For ∼5000 photons with energy at 662 keV (collected over 24 h), $MoS_2$ resonators exhibit resonance frequency shifts of ∼0.5−2.1%, corresponding to a responsivity of 30−82 Hz/photon (Figure 11.3a). The intrinsic sensitivity (limit of detection) is estimated to approach 0.02−0.05 photons, which makes these 2D resonators excellent sensors for low-dose radiation leakage that might be otherwise undetectable.[260]

*11.3.3. Inertial Sensing.* While low-dimensional materials usually have a small mass, they can form highly sensitive inertial sensors when a proof mass is attached. For example, suspended graphene membranes have been used to realize resonant accelerometers. By exploiting the piezoresistive effect in graphene, the acceleration can be measured from the resistance change of the 2D material (Figure 11.3b).[261,262] These graphene accelerometers have a device footprint at least 2 orders of magnitude smaller than conventional MEMS silicon accelerometers, which can be advantageous in terms of integration and power consumption.

*11.3.4. Pressure Sensing.* The resonance frequency of NEMS resonators can also change with pressure, which suggests potential for pressure sensing. For example, for $MoS_2$ membranes covering microtrenches, the change in environment pressure results in a pressure difference across the membrane, and $MoS_2$ membranes can exhibit different resonant responses under different pressure ranges (from ∼10 mTorr to 400 Torr (Figure 11.3c), with pressure responsivities of the resonance frequency from ∼0.05 to 0.8 MHz/Torr.[263] In another example, graphene-resonator-based squeeze-film pressure sensors have been demonstrated, which can operate between 1 and 1000 mbar with a responsivity of 9000 Hz/mbar.[264]

*11.3.5. Charge Sensing.* NEMS resonators can detect very tiny changes in force and strain induced by a charge perturbation, and thus can function as highly sensitive electrometers (charge sensors). In one example, the excellent strain sensitivity of $MoS_2$ NEMS resonators, in conjunction with the bifurcation amplifier implementation, enabled real-time detection of charges on the order of 10 electrons at room temperature.[212] Furthermore, the nonlinearity in the 2D NEMS resonators is exploited to implement a set-reset (SR) mechanical charge flip-flop which can record short-lived charge events.

## 12. RESEARCH FRONTIERS ENABLED BY NEMS RESONATORS

In addition to the rich dynamics offered by the mechanical coupling (Section 10) and the superb sensitivity to external stimuli (Section 11), 1D and 2D resonators are made of materials that can feature interesting quantum electron transport properties. With these unique features, NEMS resonators offer a plethora of intriguing possibilities for pushing the frontiers of both fundamental science and cutting-edge technology.

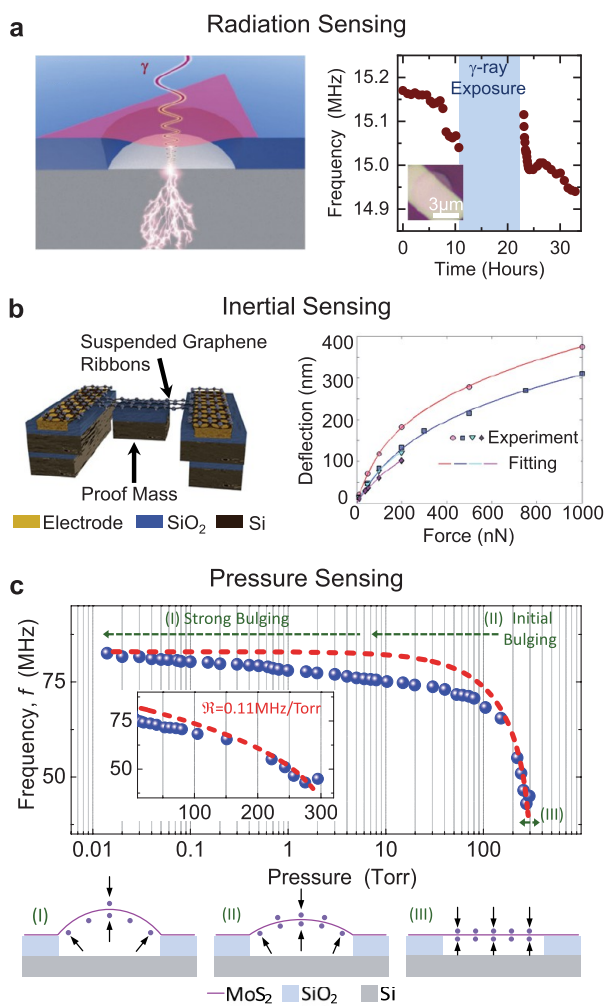

Figure 11.3. Various sensing applications demonstrated using low-dimensional NEMS resonators. (a) Schematic illustration of a 2D MoS₂ resonator used for γ-ray radiation sensing and the measured sensing data. Reprinted in part with permission from ref 260. Copyright 2016 American Institute of Physics. (b) Schematic illustration of a graphene accelerometer and sensing data. Reprinted in part with permission from ref 261. Copyright 2019 Nature Publishing Group. (c) Schematics of the cross section of a MoS₂ resonator at different pressures, and the measured dependence of the resonance frequency on chamber pressure. Reprinted in part with permission from ref 263. Copyright 2014 Institute of Electrical and Electronics Engineers.

**12.1. RF Signal Processing toward Wireless Communication.** Resonators based on low-dimensional materials may be used for RF signal processing and communication applications and could potentially exhibit performance advantages such as ultralow power consumption, high resonance frequency, and broad frequency tuning range.

The first NEMS resonator with a fundamental mode resonance frequency exceeding 1 GHz was a SiC nanobeam resonator.[265] When such an open-loop NEMS resonators is connected into a closed loop with feedback, it can exhibit sustained oscillations with only a DC supply once the circuit satisfies the Barkhausen criteria: (1) the open-loop gain equals unity, and (2) the feedback phase is an integer multiple of 2π.

Using such a construction, a SiC NW NEMS oscillator operating at 428 MHz has been demonstrated (Figure 12.1a), showing signal generation with low phase noise and excellent stability.[115] A graphene NEMS oscillator[59] (Figure 12.1b) with a 14% frequency tuning range has been realized and used as the core of a voltage-controlled oscillator (VCO), a type of device widely used in RF communication. Audio signal transmission and FM signal generation were demonstrated with this VCO. MoS₂ resonators have also been made into self-sustained feedback oscillators using an optoelectronics feedback circuitry.[266]

Besides oscillators, suspended graphene membranes have been used as RF switches, which exhibit small pull-in voltage (~1 V) and excellent signal isolation (~30 dB at 40 GHz).[267] Furthermore, using a SWCNT resonator in a field-effect transistor configuration, AM, FM, and digital demodulation has been realized with the CNT resonator as a mixer (Figure 12.1c), showing potential toward wireless communication.[70] A nanotube radio has also been demonstrated using a singly clamped resonator geometry, with the nanotube functioning simultaneously as the antenna, tunable band-pass filter, amplifier, and demodulator.[55]

**12.2. NEMS Arrays and Other Coupled Systems.** Coupling different mechanical resonators enables large-scale arrays of cascaded resonators. This can form a phononic crystal, displaying phononic bandgaps caused by mutual coupling among the resonant modes (Figure 12.2a).[222,223] In one example with graphene transferred onto an array of nanopillars, measurements suggest the formation of a phononic crystal with a quasi-continuous frequency spectrum.[268] In another example, phononic crystal waveguides in engineered h-BN nanomechanical structures have been demonstrated (Figure 12.2b), and the as-fabricated h-BN devices exhibit strong phonon dispersion relation with prominently higher first transmission band frequency and wider bandgap opening, compared with their counterparts made from conventional 3D crystalline materials with similar geometric designs.[223] Such h-BN waveguides are capable of supporting 15−24 MHz RF acoustic wave propagation over an effective length of 1.2 mm with a group velocity as high as 250 m/s at the first transmission band, while attenuating the transmission at the stop-band and bandgap by over 30 dB. Combining the unique piezoelectric properties of h-BN, such phononic structures may empower dynamically tunable devices for RF signal processing, or even toward building future integrated phononic and hybrid quantum circuitry.

In addition to phononic structures, low-dimensional NEMS resonators can also be coupled to photonic and optomechanical devices. In one example, a graphene resonator coupled to an on-chip waveguide was used to locally detect the hybridization between the propagating optical modes, by sensing changes in the optical gradient force (a kind of radiation pressure) produced by the light which affects the vibration amplitude.[269] In another example,[270] a hybrid system consisting of a SiC microdisk cavity and an h-BN nanomechanical resonator showed strong mechanical coupling, even with the thickness of h-BN scaled down to a few atomic layers. Understanding the nanomechanical properties facilitates the design and implementation of SiC/h-BN hybrid systems, toward exploring and exploiting coherent transduction between optics and mechanics on such platforms.

One interesting consequence of the coupling is that it allows for the tiny motion of a stiff bulk resonator to be amplified to a large motion in a second resonator made of a more compliant 2D material. This has been demonstrated in a graphene/silicon nitride hybrid resonant structure (Figure 12.2c).[224] In another experiment, the nonlinearity has been engineered in a silicon nitride membrane by coupling it to a graphene resonator; mechanical frequency combs have also been observed in this device, similar to the well-known optical frequency combs (Figure 12.2d).[225]

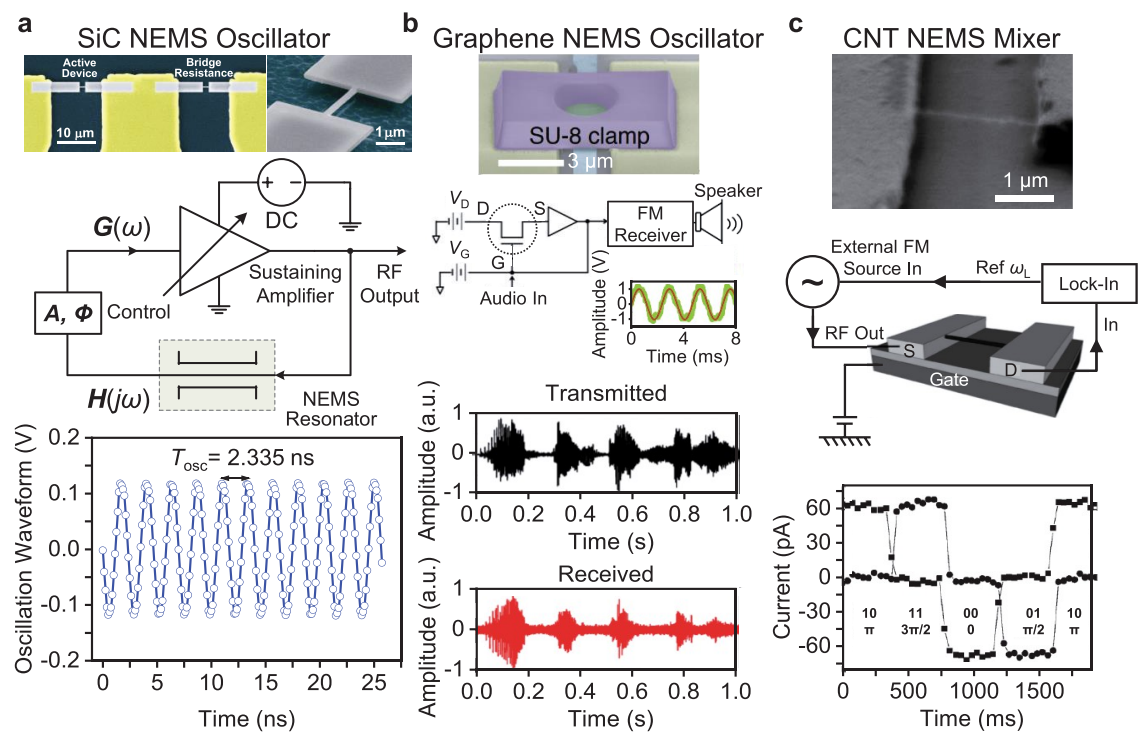

**Figure 12.1.** Low-dimensional NEMS resonators for RF signal processing. (a) False-colored SEM images of self-sustained SiC NW NEMS oscillators, the corresponding measurement circuit diagram showing a feedback loop, and the clean, stable, sinusoidal time-domain oscillation waveform of the oscillators. Reprinted in part with permission from ref 115. Copyright 2008 Nature Publishing Group. (b) False-colored SEM image of a circular graphene oscillator with SU-8 clamp and local gate electrodes, the simplified circuit for a graphene radio station, and the transmitted and received audio waveform of 1 s soundtrack from the graphene NEMS oscillator. Reprinted in part with permission from ref 59. Copyright 2013 Nature Publishing Group. (c) SEM images of a CNT resonator, FM mixing and demodulation circuit, and the measurement data showing the feasibility of digital demodulation using CNT resonator. Reprinted in part with permission from ref 70. Copyright 2010 Wiley-VCH.

**12.3. Quantum Transport in NEMS Resonators: Electromechanical Coupling.** While the mechanical motion in NEMS resonators can be coupled to different degrees of freedom, such as photons and spins, the electromechanical coupling, i.e., the interaction between mechanical vibration and electron transport properties becomes particularly interesting when the devices are measured at cryogenic temperatures.[93] Here we use CNT resonators to discuss the exotic physical processes resulting from strong electromechanical coupling.

*12.3.1. Origin of Electromechanical Coupling.* In terms of electron transport properties, CNTs are 1D wires that can become ballistic at low temperatures,[93] meaning that they can carry electrical current without noticeable dissipation along the nanotube. The electron transport through the nanotube is then dictated by the electrical transmission of the barriers formed at the interface between the nanotube and the electrical leads, giving rise to different quantum transport regimes. The carrier density of the nanotube can be tuned capacitively with the gate voltage as well as by the mechanical vibration (Figure 12.3a). The electromechanical coupling emerges through this nanotube-gate capacitance, since it depends on the nanotube displacement $z$. Mechanical vibrations of nanotubes have been coupled to electrons in different transport regimes, including single-electron tunneling,[91,96,110,229,230,271−277] Kondo,[48,278] and electronic Fabry−Pérot interference.[104]

Due to the small mass of nanotubes, the strength of the electromechanical coupling is significant. This can be understood in the following way: When adding one electron into the nanotube, the associated capacitive force $F_e$ creates a comparatively large static displacement $\Delta z = F_e/m\omega_0^2$. Since $m$ is tiny, the resulting electronic energy shift $F_e\Delta z$ is quite sizable.

Large electromechanical coupling is important for the readout of nanotube mechanical vibrations. It enables transduction of small vibration amplitudes into measurable electrical current modulations. Different methods were developed (see Section 3.2 for details) to optimize the sensitivity of the detection.[11,40,70,254,274] The noise of the detection can reach the level of the displacement noise of the zero-point fluctuations at the resonance frequency.[48] This is impressive considering that nanotubes are physically the smallest operating mechanical resonators and that the detection of mechanical vibrations becomes increasingly difficult for resonators with reduced dimensions and sizes.

*12.3.2. Effects on Mechanical Motion.* Conversely, the electromechanical coupling can also generate a large back-action of electrons on the nanotube mechanical vibrations. This back-action has been used to tune the resonance frequency, the energy decay time, and the effective temperature of the resonator.[48,276] Devices based on single-electron tunneling are of main importance in this context, since the back-action of electrons in this case is comparatively much larger than in other electron transport regimes.

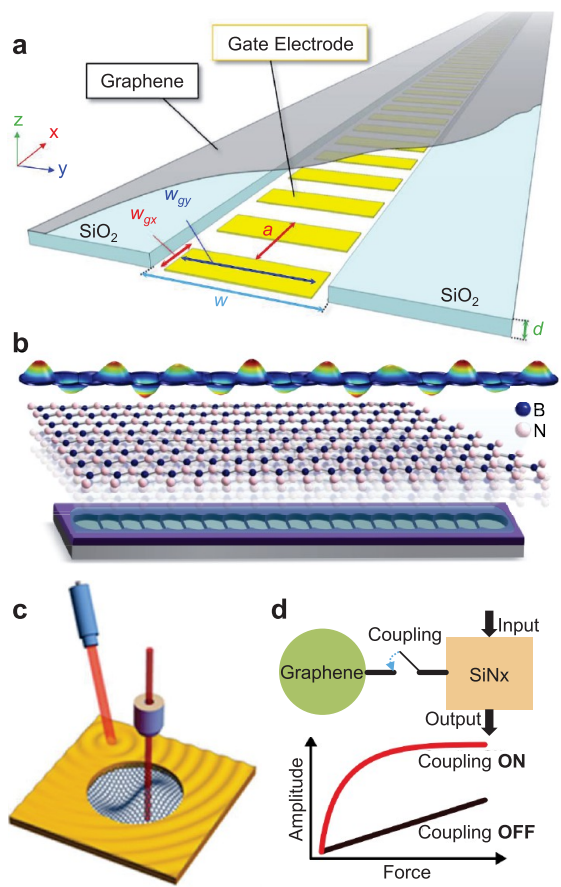

**Figure 12.2. Advanced coupled NEMS devices.** (a) Schematic of a graphene-based acoustic waveguide, where the graphene film is suspended over a trench with a gate electrode array at its bottom. Reprinted with permission from ref 222. Copyright 2019 American Physical Society. (b) Schematic of an h-BN phononic crystal waveguide. Reprinted in part with permission from ref 223. Copyright 2019 American Chemical Society. (c) Schematic of a graphene/silicon nitride hybrid resonant structure. Reprinted in part with permission from ref 224. Copyright 2020 American Chemical Society. (d) Giant tunable mechanical nonlinearity in a graphene−silicon nitride hybrid resonator. Reprinted in part with permission from ref 225. Copyright 2020 American Chemical Society.

Single-electron tunneling in a nanotube quantum dot emerges when the transmission of the tunnel barriers is sufficiently low (Figure 12.3a).[93] When an electron tunnels into the nanotube dot, it creates a Coulomb gap that prevents a second electron from tunneling into the nanotube dot, i.e., "Coulomb blockade". After some time, the extra electron on the nanotube dot tunnels out of it. As a result, the system fluctuates back and forth between the two states with $N$ and $N + 1$ electrons. In an experiment,[271] the electrical transport of the nanotube quantum dot features a series of conductance peaks when sweeping the gate voltage, see Figure 12.3b, top. The peaks are associated with situations where the energies of the electron states $N$ and $N + 1$ are degenerate, which results in rapid fluctuations. In between the peaks, these fluctuations can be suppressed almost to zero ("Coulomb blockade" regions), such that the average charge number on the quantum dot is a well-defined integer.

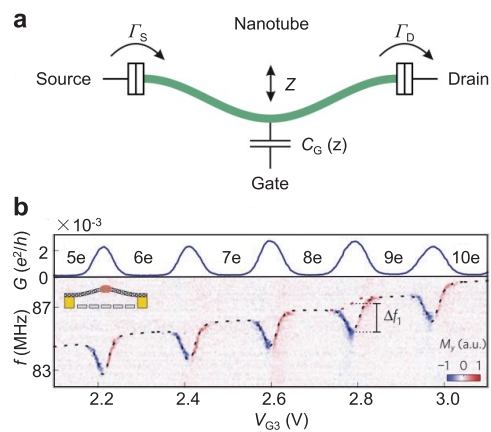

**Figure 12.3. CNT resonator as a quantum dot.** (a) Sketch of a CNT coupled to source and drain leads with charge transport rates $\Gamma_{S,D}$. The total electrical charge of the nanotube can be tuned capacitively by a gate voltage. The nanotube-gate capacitance $C_G(z)$ depends on the nanotube displacement $z$. (b) Electrical conductance and mechanical resonance frequency as a function of the gate voltage measured at a temperature of 16 K. Reprinted in part with permission from ref 272. Copyright 2014 Nature Publishing Group.

The charge fluctuations in the single-electron tunneling regime result in a reduction of the resonance frequency (Figure 12.3b, bottom). The size of this frequency reduction is a direct measure of the electromechanical coupling strength. It was shown that the single-phonon single-electron electromechanical coupling strength $g_0 = F_e/\hbar z_{zp}$ can become comparable to the mechanical resonance frequency $\omega_0$,[277] where $z_{zp}$ is the zero-point motion amplitude of the vibrations. This so-called ultrastrong coupling regime is analogous to those achieved in electro- and optomechanical devices,[279−281] and various phenomena may emerge in this extreme regime.

In addition to the reduction of the resonance frequency, the single-electron tunneling back-action can increase by a large amount the mechanical losses,[96,110] the Duffing nonlinearity (Section 9.1),[273] and the dispersive coupling between the mechanical modes of the nanotube,[238] (Section 10.1). Some of these back-action effects have been observed in double quantum dots where the Coulomb interaction between electrons also plays a central role.[271,274]

Similar effects of electron transport on mechanical resonances have also been observed in graphene resonators. By combining magnetotransport and resonance measurements, researchers have demonstrated periodic modulations of the resonance frequency using gate voltage and magnetic field, which tune the electron array through the discrete Landau levels (Figure 12.4).[282] Based on such electron-magneto modulation of resonance frequencies, researchers are able to determine the internal chemical potential oscillations in the 2D atomic layers.

**12.4. Putting Mechanics into Quantum Mechanics.** Nowadays, one key theme of the "second quantum revolution" is *quantum engineering*, which is to gain full understanding and control of man-made quantum objects that generate, store, transmit, and process quantum information. NEMS can be used for probing quantum-limited phenomena, for realizing quantum bits (qubits), or as NEMS or phononic transducers for qubits.[283] Previous results, for example, include coupling acoustic resonators to superconducting qubits,[284,285] using superconducting qubits to perform quantum-limited measurements of a bulk dilatational resonator,[286] and observing quantum entanglement[287,288] in two mechanical drumhead resonators.[234] These examples can provide important guidelines for using low-dimensional NEMS as qubits or in qubit-coupled systems.

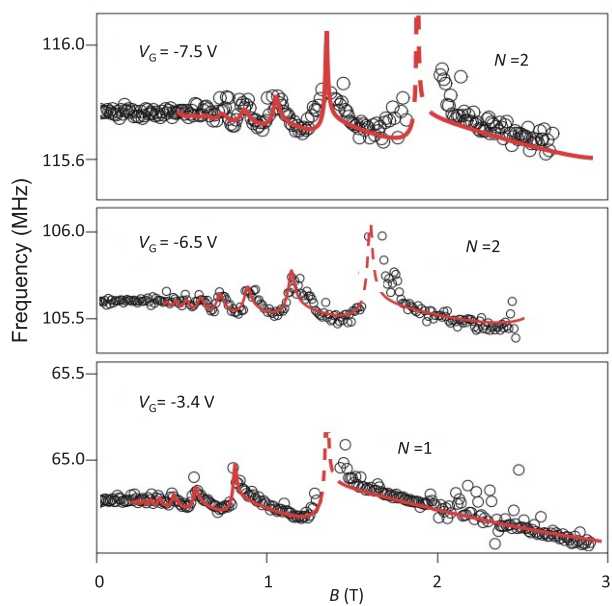

**Figure 12.4.** Mechanical resonant frequency in a graphene resonator as functions of the applied magnetic fields at different $V_G$ and corresponding fits (red curves). Reprinted in part with permission from ref 282. Copyright 2016 Nature Publishing Group.

Extensive efforts have been devoted to reaching the quantum ground state (QGS) and more complex nonclassical states with low-dimensional NEMS devices. The QGS corresponds to the case of negligible classical (thermal) fluctuations, where the dynamics of the resonator is dominated by its zero-point fluctuations, as required by the Heisenberg uncertainty relation.[289] This state is characterized by an average phonon number $n \ll 1$. At high temperature, the phonon number can be estimated as $n \approx k_B T / h f_0$, with Boltzmann's constant $k_B$, the temperature $T$, Planck's constant $h$, and the resonance frequency $f_0$ (note that $h f_0 = \hbar \omega_0$ is the energy of a single phonon of such frequency). The QGS can be reached for high frequencies and low temperatures, using for instance a resonator with $f_0 = 10$ GHz in a dilution refrigerator with $T = 10$ mK. However, many NEMS have frequencies in the 1−100 MHz range and cannot be cooled to the QGS passively.

Where passive cooling is insufficient, alternative cooling schemes have been proposed or implemented in NEMS resonators based on 1D and 2D nanomaterials, such as sideband ground-state cooling of a graphene resonator[77,79] or photothermal back-action cooling of graphene resonators,[86] $MoS_2$ resonators,[266] and CNT resonators,[80] offering encouraging prospects on this research front.

*12.4.1. Sideband Cooling.* Among the cooling schemes that have been successfully demonstrated, parametric sideband cooling, as extensively explored in cavity optomechanics,[292] has proven to be particularly useful. Sideband cooling into the QGS is routinely achieved with top-down microfabricated resonators coupled to either optical cavities or superconducting resonators.[235,290,291] Graphene flakes have been coupled to superconducting resonators,[78,79,125] and sideband cooling has been used to achieve a thermal population with $n_m = 7.2 \pm 0.2$.[77] Sideband cooling can also be achieved by replacing the

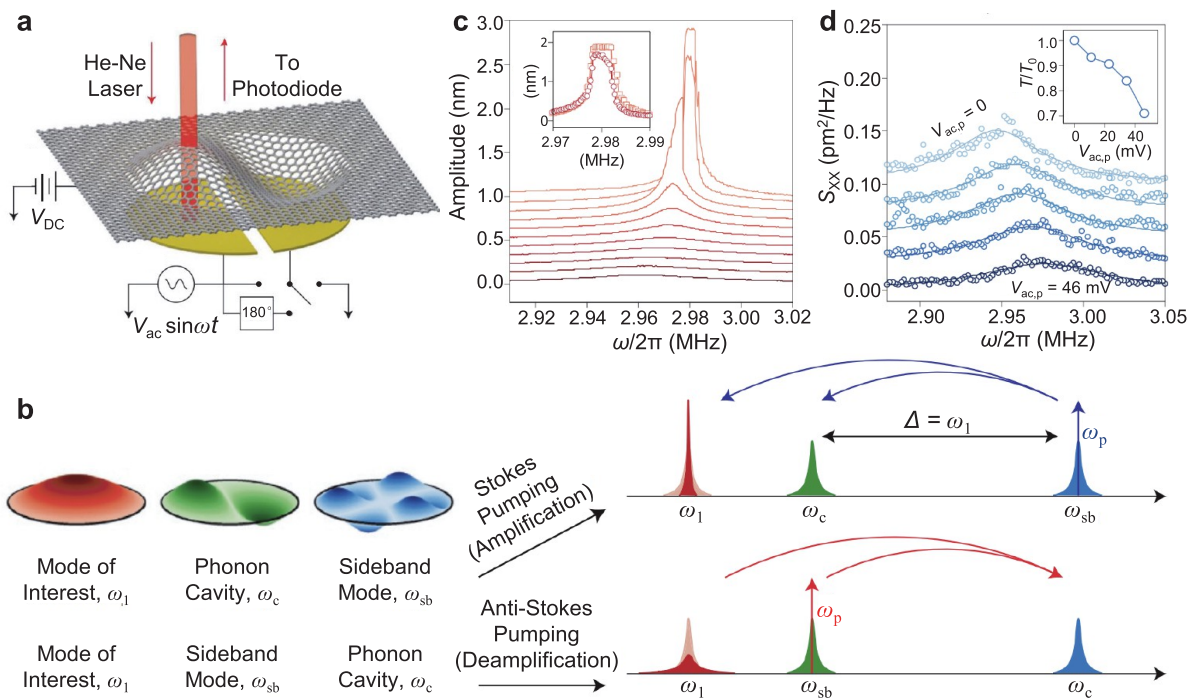

**Figure 12.5.** Parametric sideband cooling in a graphene resonator. (a) Schematic of the experimental setup. (b) Schematic of sideband pumping in frequency space. The curved arrows indicate the direction of energy flow when the system is pumped at $\omega_p$. (c) Amplification of a mode as another mode is pumped at its Stokes sideband. (d) Parametric cooling in a mode on pumping the anti-Stokes sideband of another mode. (a−d) Reprinted in part with permission from ref 228. Copyright 2016 Nature Publishing Group.

superconducting resonator by a high-frequency mechanical mode (Figure 12.5).[228]

As previously discussed (see Section 10.2), the presence of a pump at the frequency difference of two modes (or resonators), $\omega_p = \omega_2 - \omega_1$, can enable efficient energy exchange between them. We previously discussed the strong coupling limit (Section 10), where the energy exchange between the modes is faster than their combined energy decay rate $\Gamma_1 + \Gamma_2$ and the modes hybridize (eq 10.2). For sideband cooling, we are interested in a different configuration.

Let us assume that mode 2 is at a sufficiently high frequency to ensure $n \ll 1$ and that its dissipation rate is higher than the parametric energy exchange rate. When parametric coupling is activated, a low-frequency mode 1 (with $n \gg 1$ in thermal equilibrium) exchanges energy with mode 2. Energy flows predominantly from mode 1 to mode 2, where it is dissipated to the environment before it can return to mode 1. Through this interaction, the thermal population of mode 1 is reduced below its thermal equilibrium value, i.e., mode 1 is cooled. The method is known as "red sideband cooling", referring to the pump frequency $\omega_p$ as a sideband detuned from $\omega_2$ by the amount $\omega_1$ on the negative (red) side. Such sideband cooling has been rather modest so far, since the resonance frequency $\omega_2$ of mode 2 has not been sufficiently large compared to the resonance frequency $\omega_1$ of mode 1, and since the lowest temperature that mode 1 can reach is given by $T_{\text{eff}} = T\omega_1/\omega_2$.[190] For positive (blue) detuning $\omega_p = \omega_2 + \omega_1$, the number of phonons is increased in both modes, i.e., the two modes "heat up". The parametric pumping can be further enhanced when a mechanical mode is present at the pumping frequency $\omega_p$.[228]

*12.4.2. Cooling through Electromechanical Coupling.* Another effective approach is to actively "cool down" the resonant motion via electromechanical coupling (Section 12.2), which has been extensively explored in CNT resonators due to their superb electron transport properties. In the most extreme cases, the electromechanical coupling can dramatically modify the charge stability diagram measurements of nanotube quantum dots.[96] These measurements indicate situations where the electromechanical system experiences instability and generates mechanical self-oscillations.[48,276] More specifically, the DC power injected into the nanotube (by applying a DC voltage bias between the source and drain electrodes) fuels large mechanical vibrations at the resonance frequency. The self-oscillations arise from the electron back-action, which results in a mechanical dissipation rate that is effectively zero or negative. The back-action origin can be single-electron tunneling[276] as well as an electrothermal effect.[48] Conversely, the back-action can also increase the total dissipation rate, such that the DC power injected into the nanotube cools the mechanical vibrations. Cooling to an average number of quanta $n_m = 4.6 \pm 2.0$ has been achieved in this way.[48]

In some cases, it is interesting to suppress the back-action associated with single-electron tunneling by operating the system in the Kondo and the electronic Fabry−Pérot regimes. This is especially relevant to boosting the $Q$-factor, since the electronic contribution to mechanical losses is then minimized.[48,104] Furthermore, the electromechanical coupling is a powerful tool to increase the coupling between the mechanical vibrations and photons. Single-electron tunneling has recently been used to amplify the coupling between the mechanical vibrations of a nanotube and the photons of a superconducting resonator.[275] This is encouraging for future research and applications, since superconducting resonators enable efficient

detection and cooling of mechanical vibrations of top-down fabricated resonators.[292]

*12.4.3. Proposals of NEMS Qubits.* A recent theory proposal showed that under particular conditions, the electromechanically induced nonlinearity of a nanotube resonator can become large enough to create a mechanical qubit.[293] By coupling the mechanical vibrations to a double quantum dot hosted along a nanotube and using realistic device parameters, it should be possible to reach the regime where the energy levels of the resonator are no longer evenly spaced. The two lowest energy levels can then be individually addressed with a resonant drive and form a qubit. This regime can be reached in the ultrastrong coupling limit when $g_0 > \omega_0$. Similar schemes have also been proposed with other NEMS layouts.[294,295] Such a device may offer opportunities for storing and processing quantum information in these man-made atomic-scale devices.

# 13. SUMMARY AND OUTLOOK: RESONANT WITH A VIBRANT FUTURE

## 13.1. Summary.
As we have presented in this review, the study of low-dimensional resonators has come a long way and is still progressing rapidly. Below we summarize some of the key messages from several sections.

*13.1.1. Material and Fabrication.* Many types of 1D nanomaterials, such as CNT and different NWs, have been studied for NEMS resonators. 2D materials offer an even greater choice: from graphene and 2D semiconductors to wide-bandgap and magnetic 2D materials, and all these nanomaterials have been explored in NEMS studies. Furthermore, given the rapidly growing 2D material family and the nearly endless possibilities of creating vdW HSs, there are many more yet to be explored. In terms of fabrication techniques, both lithography/etching and transfer approaches have been used in making NEMS resonators. The former starts with an unsuspended (supported) device structure before releasing the nanomaterial, which is widely used in making doubly clamped structures. The latter is often used for producing fully clamped resonators: it starts with substrates with prepatterned structures and then places the nanomaterial over such a patterned area, directly forming suspended devices.

*13.1.2. Measurement Schemes.* Due to the much smaller device size and motion amplitude compared with most MEMS devices, different measurement techniques need to be devised specifically for these NEMS resonators. A number of experimental schemes have been developed and successfully demonstrated for NEMS resonators. Among them, electrical readout and optical readout are the most widely used ones. The former is commonly used for 1D devices and doubly clamped 2D ones, given that all the current flowing through the device carries the motional signal; the latter is widely used for fully clamped 2D resonators, whose motional part is typically larger than the laser spot.

*13.1.3. Tuning Resonance Frequency and Quality Factor.* Given the high flexibility of low-dimensional nanomaterials, the device resonance frequency can be effectively tuned through strain, which can be induced through external electric fields or other schemes. Frequency tuning of more than 1000% has been achieved in both 1D and 2D NEMS resonators, greatly exceeding that in most MEMS resonators. Quality factors in NEMS resonators can strongly depend on a number of factors, in particular temperature. At cryogenic temperatures, $Q$ values over 1 million have been achieved in both 1D and 2D resonators.

*13.1.4. Nonlinear Responses.* In NEMS resonators, due to the small device size and their low stiffness, the vibrational amplitude can easily become much larger than the device dimension in the direction of motion, which can lead to strong nonlinear effects. Thermal fluctuation is also more pronounced in these minuscule devices. Despite strong nonlinear effects and significant thermal noise that both limit the linear DR, broad DR (over 70 dB) has been achieved in both 1D and 2D NEMS resonators.

*13.1.5. Coupling in the Mechanical Domain.* Coupling between two resonators or between two resonant modes within the same resonator has been demonstrated in both 1D and 2D NEMS devices. Such coupling can occur through nonlinear effects or parametric pumping, with the latter requiring an external signal at a third frequency (which is often the sum or the difference of the resonance frequency of the two modes). The coupling effects can lead to rich physics in these nanodevices, in particular the controlled transfer of energy between different modes.

*13.1.6. Device Applications.* These minuscule devices are highly sensitive to external stimuli, making them particularly suitable for sensing applications. Both yg mass resolution and $zN/\sqrt{Hz}$ force sensitivity have been achieved using CNT resonators at cryogenic temperature. Other sensing functions have also been demonstrated, including temperature, light and radiation, acceleration, pressure, and electric charge. Besides sensing, they have also been used for RF signal processing as well as fundamental studies related to quantum transport and quantum information processing.

**13.2. Challenges toward Scaling.** While many encouraging milestones have been achieved, there are still some important challenges. Several of them are related to fabrication and scaling, which we discuss below.

*13.2.1. Larger-Scale Fabrication of Functional Devices.* One advantage of these devices is their smaller device footprint compared with most MEMS devices, which could lead to denser integration. However, it is still challenging to fabricate these suspended NEMS devices at large scale. While some early attempts had led to encouraging results,[60,62] the yield is still not ideal for practical applications. Nevertheless, with the fast advancement of large-scale growth methods[296–303] and clean transfer techniques,[304–309] which involve a lot of research progress in industry and semiconductor foundries, we envision that this challenge will be addressed as the entire field progresses.

*13.2.2. Better Predictability and Uniformity of Device Performance.* As the production of large-scale device arrays becomes increasingly reproducible, it will be important to look at the predictability of device performance, which is critical for designing and realizing functional devices on a wafer scale. However, the device performance can be affected by a number of factors such as fabrication-induced tension,[17,19,20,310] wrinkles in the suspended region,[311–313] contaminations on the surface,[57] and slight variation in device dimension.[170] While these are difficult to completely avoid given the current state of nanofabrication technology, the development of annealing techniques,[45,48,57,61] control of the environment during fabrication,[304,307] improved transfer techniques,[314,315] and even transfer-free approaches[316,317] are promising when the process is standardized in foundries. Another possible solution is to design the device such that the device performance (such as frequency) is insensitive to these factors. For example, by designing devices operating in the appropriate mechanical regime,[310] one can minimize the effect of tension and thus produce devices with highly predictable resonant frequencies.

*13.2.3. More Scalable Readout Scheme.* As the level of device integration increases, compatible readout schemes are required. While optical readout has proven to be highly versatile and effective in fundamental research, for practical applications, electrical schemes are much more compatible with highly integrated NEMS devices. However, existing electrical readout techniques typically require external circuity and additional instruments, which poses challenges for using NEMS devices in standalone wireless nodes, or as logical components in highly integrated circuits. Toward such goals, one possible solution is on-chip functional circuity such as phase-locked loop. However, such circuity requires the performance of the NEMS resonator to be highly predictable, as outlined in the point above, so that the electrical components in the circuit can be properly designed.

*13.2.4. Further Miniaturization of Device Structures.* While current NEMS resonators can be atomically thin, they are typically micrometer-sized in at least one of the lateral dimensions. Toward higher frequency and integration, the lateral device size needs to be further scaled down, which poses challenges for fabrication. Furthermore, smaller devices lead to smaller motional signal. Therefore, a number of technical challenges need to be addressed toward this goal.

**13.3. Future Opportunities.** Moving forward, a number of applications could emerge by leveraging the unique properties of low-dimensional NEMS resonators. Here we discuss a few more possible directions for future research in this field as well as the technological challenges that need to be addressed in achieving future goals.

*13.3.1. Higher-End and Further Miniaturized Sensors.* With the superb sensitivity and responsivity as well as the miniscule device footprint and ultralow power consumption, 1D and 2D NEMS resonators have the potential to enable advanced sensor technologies. For instance, the possibility of measuring the tiny (zeptonewton) forces generated by individual nuclear spins can lead toward nanoscale magnetic resonance imaging, i.e. NanoMRI, and hence the direct investigation of complex molecules and their functionality.[256] Similarly, the ability to measure the mass of a single atom could have transformative consequences for the field of proteomics, as it would allow rapid and precise characterization of large proteins without the need of ionizing the protein.[246,336,318]

To date, however, the demonstration of sensitivities required to detect individual atoms[250] or zeptonewton forces[91] have only been demonstrated in cryogenic temperatures under high vacuum. Translating these excellent device characteristics into practical applications is a tremendous challenge for a variety of reasons.

First, allowing a pristine nanomechanical resonator to interact with a sample can significantly deteriorate its sensitivity. A famous example of such an effect is the notorious noncontact friction between closely spaced bodies, which limits the sensitivity of some scanning force microscopy experiments.[319–327] In general, 1D and 2D resonators are even more sensitive to such perturbations due to their low masses and reduced spring constants.[104] Overcoming this challenge will be possible once we have a better understanding and control of surfaces on the atomic scale. Interestingly, we can make use of the sensitivity of nanomechanical resonators to investigate their own surface properties and then employ this knowledge to improve their performance. This line of future work is very much aligned with ongoing efforts in the trapped-ion and super-conducting qubit communities, whose devices are likewise limited by surface effects.[328−335] NEMS sensors could therefore act as a catalyst for the progress in various fields of modern physics.

A second challenge for practical sensing applications, as described in Section 13.2.2, is the uniformity of resonator properties. As features decrease toward the atomic scale, the influence of individual material defects increases. This will make it hard to produce sensors with reproducible and scalable properties.[170] As previously discussed, scalability is important, for instance, for producing resonator grids for enhanced mass-sensing throughput. This issue must be solved by making use of the intrinsic precision of bottom-up processes; if materials are grown from well-defined starting conditions with low enough defects, the resulting devices are highly reproducible without the need of atomic-scale manipulation.[60]

A third important step toward practical usage will be to enable room-temperature operation of sensors with adequate performance. Avoiding the costs and work hours required for cryogenic experiments will bring devices much closer to marketable applications. Operating a mechanical sensor at higher temperatures usually degrades its performance due to the increased thermomechanical force noise (eq 11.2). This drawback, however, can be mitigated by using resonators with low effective mass and high quality factors. For instance, a CNT cantilever at 300 K can have a force sensitivity below 1 aN per root Hz, as shown in ref 80, better than what most larger state-of-the-art sensors can achieve at cryogenic temperatures.[253,255] It may be rewarding to achieve adequate performance in these exquisite machines in close to ambient conditions or at conditions achievable with on-chip technology (such as using thermo-electric cooling or MEMS vacuum pumps). However, how to translate such excellent device performance into practical applications has remained an important question[246,336] and will likely require extensive engineering efforts.

### 13.3.2. Logical Processing Units.

Resonator-based logic gates[337−340] have been extensively explored to realize computing paradigms beyond the von Neumann architecture.[341] NEMS resonators based on low-dimensional nanomaterials are particularly promising for realizing frequency-shift-based logic functions, as the superb mechanical properties allow the device frequency to be broadly and relatively easily changed. Together with the ultrasmall device volume, 1D and 2D NEMS resonators hold promise for ultralow power logic operations. Toward this vision, one important hurdle to overcome is the large-scale production of highly uniform resonator devices[60,62] (ideally compatible with CMOS back-end-of-line (BEOL) processes,[61,342] as illustrated in Figure 13.1), which is a prerequisite for the scaling of such technologies.[343]

One particular form of computing that has spurred a lot of interest in recent years is the neuromorphic solving of hard optimization problems with Hopfield networks.[344−346] Strong parametric pumping can turn a nonlinear resonator into a bistable system that fulfils the role of an artificial neuron. Networks of such parametric oscillators, or "parametrons", can then be used to simulate the behavior of a problem and find an optimal solution through, e.g., quantum annealing.[347−351] 1D and 2D NEMS could become a valuable resource for such networks due to their low power consumption, the possibility of dense packaging, and their simple integration with electrical components.[98,114] As for other applications, the main experimental challenges lie in the fabrication of nearly identical resonators with highly uniform properties. Addressing these issues is therefore crucial for a wide range of potential applications and should be a priority for the community in the coming years. In addition, it will be important to understand the complexity of nonlinear oscillator networks, which can go beyond that of an Ising Hamiltonian.[350,352]

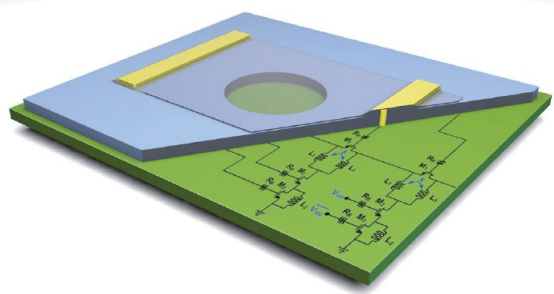

**Figure 13.1.** Artistic illustration of a 2D NEMS resonator integrated with a CMOS circuit.

### 13.3.3. RF Components toward 6G Technology and Beyond.

MEMS resonator technology has been extensively used in wireless communications. The scaling law of NEMS resonators[17,353,354] suggests that these truly nanoscale devices have the potential to operate in the GHz bands, and in CNT resonators, the resonance frequency could be pushed above 10 GHz.[103,102] By exploiting certain resonant modes,[355] these resonators might even be useful for communications in millimeter-wave or THz bands. In addition, the excellent frequency tunability[63] and reconfigurability[182] in NEMS resonators make them particularly promising for multiband RF operations, which are important for 5G and 6G communications where many different electromagnetic frequency bands are used. A major challenge is the impedance mismatch with silicon-based circuits, which can affect the insertion loss of NEMS devices. This is particularly important as frequency increases, as parasitic effects become more pronounced. Therefore, it is important for researchers to discover and develop effective measures to address this issue before 1D and 2D NEMS resonators can be used as RF circuit components.

We close by saluting Richard Feynman's vision that "there is plenty of room at the bottom". With the encouraging achievements over the last two decades and the promising opportunities in front of us, there has never been a more exciting time to be engaged in the scientific and technological exploration of these truly nanoscale man-made machines, which are undoubtedly resonant with a vibrant future.

## AUTHOR INFORMATION

### Corresponding Authors

**Akshay Naik** − Centre for Nano Science and Engineering, Indian Institute of Science, Bangalore 560012 Karnataka, India; orcid.org/0000-0001-6325-7231; Email: anaik@iisc.ac.in

**Rui Yang** − University of Michigan−Shanghai Jiao Tong University Joint Institute, Shanghai Jiao Tong University, Shanghai 200240, China; School of Electronic Information and Electrical Engineering, Shanghai Jiao Tong University, Shanghai 200240, China; orcid.org/0000-0002-6163-2904; Email: rui.yang@sjtu.edu.cn

**Philip X.-L. Feng** − Department of Electrical and Computer Engineering, Herbert Wertheim College of Engineering, University of Florida, Gainesville, Florida 32611, United

States; ● orcid.org/0000-0002-1083-2391;
Email: philip.feng@ufl.edu

**Zenghui Wang** − *Institute of Fundamental and Frontier Sciences, University of Electronic Science and Technology of China, Chengdu 610054, China; State Key Laboratory of Electronic Thin Films and Integrated Devices, University of Electronic Science and Technology of China, Chengdu 610054, China;* ● orcid.org/0000-0003-3743-7567;
Email: zenghui.wang@uestc.edu.cn

## Authors

**Bo Xu** − *Institute of Fundamental and Frontier Sciences, University of Electronic Science and Technology of China, Chengdu 610054, China;* ● orcid.org/0000-0003-0262-2017

**Pengcheng Zhang** − *University of Michigan−Shanghai Jiao Tong University Joint Institute, Shanghai Jiao Tong University, Shanghai 200240, China;* ● orcid.org/0000-0003-4106-6762

**Jiankai Zhu** − *Institute of Fundamental and Frontier Sciences, University of Electronic Science and Technology of China, Chengdu 610054, China;* ● orcid.org/0000-0002-5495-7612

**Zuheng Liu** − *University of Michigan−Shanghai Jiao Tong University Joint Institute, Shanghai Jiao Tong University, Shanghai 200240, China;* ● orcid.org/0000-0002-1866-2649

**Alexander Eichler** − *Department of Physics, ETH Zurich, 8093 Zurich, Switzerland;* ● orcid.org/0000-0001-6757-3442

**Xu-Qian Zheng** − *Department of Electrical and Computer Engineering, Herbert Wertheim College of Engineering, University of Florida, Gainesville, Florida 32611, United States; College of Integrated Circuit Science and Engineering, Nanjing University of Posts and Telecommunications, Nanjing 210023, China;* ● orcid.org/0000-0003-4705-771X

**Jaesung Lee** − *Department of Electrical and Computer Engineering, Herbert Wertheim College of Engineering, University of Florida, Gainesville, Florida 32611, United States; Department of Electrical and Computer Engineering, University of Texas at El Paso, El Paso, Texas 79968, United States;* ● orcid.org/0000-0003-0492-2478

**Aneesh Dash** − *Centre for Nano Science and Engineering, Indian Institute of Science, Bangalore 560012 Karnataka, India;* ● orcid.org/0000-0002-6465-6506

**Swapnil More** − *Centre for Nano Science and Engineering, Indian Institute of Science, Bangalore 560012 Karnataka, India;* ● orcid.org/0000-0001-7906-8902

**Song Wu** − *Institute of Fundamental and Frontier Sciences, University of Electronic Science and Technology of China, Chengdu 610054, China;* ● orcid.org/0000-0002-3104-1855

**Yanan Wang** − *Department of Electrical and Computer Engineering, Herbert Wertheim College of Engineering, University of Florida, Gainesville, Florida 32611, United States; Department of Electrical and Computer Engineering, University of Nebraska-Lincoln, Lincoln, Nebraska 68588, United States;* ● orcid.org/0000-0002-9663-4491

**Hao Jia** − *Shanghai Institute of Microsystem and Information Technology, Chinese Academy of Sciences, Shanghai 200050, China;* ● orcid.org/0000-0002-1429-6995

**Adrian Bachtold** − *ICFO-Institut de Ciencies Fotoniques, The Barcelona Institute of Science and Technology, Barcelona 08860, Spain;* ● orcid.org/0000-0002-6145-2479

Complete contact information is available at:
https://pubs.acs.org/10.1021/acsnano.2c01673

## Author Contributions

[†]These authors contributed equally to this work.

## Notes

The authors declare no competing financial interest.

## ACKNOWLEDGMENTS

We gratefully acknowledge support from National Natural Science Foundation of China (grants U21A20459, U21A20505, 62150052, 62104029, 12104086, 62004026, 62004032, 62104140, 62104241), Sichuan Science and Technology Program (grants 2021YJ0517, 2021JDTD0028), Fundamental Research Funds for the Central Universities (ZYGX2020ZB014 and ZYGX2020J029), Lingang Laboratory Open Research Fund (grant LG-QS-202202-11), Science and Technology Commission of Shanghai Municipality (STCSM) Natural Science Project General Program (grant 21ZR1433800), Shanghai Sailing Program (grant 19YF1424900), and Shanghai Pujiang Program (grant 20PJ1415600). P.F. acknowledges National Science Foundation (NSF) CAREER Award (grant no. ECCS-1454570, ECCS-2015708), CCSS Program (grant no. ECCS-1509720), EPMD Program (grant no. ECCS-1810154, ECCS-2015670), and EFRI Program (grant no. EFMA-1641099). A.B. acknowledges ERC Advanced grant no. 692876, AGAUR (grant no. 2017SGR1664), MICINN grant no. RTI2018-097953-B-I00, QUANTERA grant (PCI2022-132951), the Fondo Europeo de Desarrollo, the Spanish Ministry of Economy and Competitiveness through Quantum CCAA and CEX2019-000910-S [MCIN/AEI/10.13039/501100011033], Fundacio Cellex, Fundacio Mir-Puig, and Generalitat de Catalunya through CERCA. A.D., S.M., and A.N. acknowledge funding support from MHRD, MeitY, and DST Nano Mission through NNetRA. Y.W. gratefully acknowledges the financial support from the National Science Foundation (NSF) via Nebraska's NSF EPSCoR RII Track-1 project and Emergent Quantum Materials and Technologies (EQUATE, OIA-2044049). We thank Dr. Changyao Chen, Dr. Yuehang Xu, Dr. Fan Ye, Dr. Xuge Fan, Dr. Yong Xie, and Dr. Shengwei Jiang for helpful discussions as well as high-resolution figures.

## 📖 Recommended by ACS